\documentclass[twocolumn,twocolappendix]{aastex7}

\usepackage{amsmath}
\usepackage{siunitx}
\usepackage{bm}
\usepackage{bbm}

\usepackage{layouts}
\usepackage{printlen}

\DeclareMathOperator*{\argmin}{arg\,min}

\begin{document}

\title{Robust CMB B-mode analysis with Needlet-ILC and simulation-based inference}

\author[orcid=0000-0003-2856-2382,sname='au1']{Adriaan J.\ Duivenvoorden}
\affiliation{Max-Planck-Institut f\"{u}r Astrophysik, Karl-Schwarzschild-Str.\ 1, 85748 Garching, Germany}
\email[show]{adriaand@mpa-garching.mpg.de}  

\author[orcid=0000-0002-7611-6179,sname='au2']{Kristen Surrao}
\affiliation{Department of Physics, Columbia University, New York, NY 10027, USA}
\email{}  

\author[orcid=0000-0002-3568-3900,sname='au3']{Adrian E.\ Bayer}
\affiliation{Center for Computational Astrophysics, Flatiron Institute, 162 5th Avenue, New York, NY 10010, USA}
\affiliation{Department of Astrophysical Sciences, Princeton University, Peyton Hall, Princeton, NJ 08544, USA}
\email{}  

\author[orcid=0000-0002-5736-5524,sname='au4']{Alexandre E.\ Adler}
\affiliation{Department of Physics, University of California Berkeley, Berkeley, CA, USA}
\affiliation{Lawrence Berkeley National Laboratory, Berkeley, CA, USA}
\email{}  

\author[orcid=0009-0006-7382-1434,sname='au5']{Nadia Dachlythra}
\affiliation{Department of Physics, University of Milano-Bicocca, Piazza della Scienza 3, 20126, Milano, Italy}
\email{}  

\author[orcid=0009-0006-7382-1434,sname='au6']{Susanna Azzoni}
\affiliation{Department of Physics, Princeton University, Jadwin Hall, Princeton, NJ 08544, USA}
\email{}  

\author[orcid=0000-0002-9539-0835,sname='au7']{J.\ Colin Hill}
 \affiliation{Department of Physics, Columbia University, New York, NY 10027, USA}
 \email{}

\begin{abstract}

We explore a novel analysis framework for parameter inference with large-scale CMB polarization data. Our method uses simulation-based inference combined with the needlet internal linear combination (NILC) algorithm and cross-correlation-based statistics to compress the data into a vector that is robust to model misspecification and small enough to be amenable to neural posterior estimation with normalizing flows. 
By leveraging this compressed data representation, our method enables the robust use of the anisotropic and non-Gaussian information in the foreground fields to more accurately separate the CMB polarization signal from these contaminants. Using an idealized ground-based experimental setup inspired by the Simons Observatory Small Aperture Telescopes, we demonstrate improved statistical constraining power for the tensor-to-scalar ratio $r$ compared to the (constrained) NILC algorithm and improved robustness to complex foregrounds compared to other techniques in the literature. Trained on a relatively simple semi-analytical foreground model, the method yields unbiased $r$ results across a range of PySM Galactic foreground simulations, including the high-complexity \texttt{d12} model, for which we obtain $r=(1.09 \pm 0.27)\cdot 10^{-2}$ for input $r=0.01$ and sky fraction $f_{\mathrm{sky}} = 0.21$.  We thus demonstrate the feasibility and advantages of a complete, maps-to-parameters, simulation-based analysis of large-scale CMB polarization for current ground-based observatories.

\end{abstract}



\section{Introduction} 

One of the main goals of current  cosmic microwave background (CMB) observatories is the search for an imprint of a background of primordial gravitational waves in the large-scale $B$-mode polarization pattern in the CMB, as quantified by the tensor-to-scalar ratio $r$~\citep{so_forecast_2019,cmbs4_science}. Galactic  emission, in the form of diffuse polarized emission from interstellar dust at $\gtrsim \SI{70}{\giga \hertz}$ and synchrotron emission at lower frequencies, dominates the polarized microwave sky on the large angular scales where the primordial signal is expected to be observable \citep{page_2007, planck_2015_x_diffuse}. The ability to robustly constrain a primordial signal will be strongly influenced by the fidelity of Galactic foreground modeling and subtraction, which is expected to be among the leading sources of uncertainty.

It is understood that the spectral energy distribution
(SED) of the polarized dust and synchrotron emission varies over the sky and along the line of sight \citep{planck_xxii_2015,planck_L_2017,pelgrims_2021}. A priori, it is difficult to model these variations accurately and current data are still too noisy to provide a sufficiently detailed description at the $r$ values targeted by upcoming experiments. The SED variations, when unmodeled, partly spoil our ability to remove the foreground contamination by extrapolating the observed foreground properties across frequencies: they ``decorrelate'' the foreground fields.
While current cosmological constraints from $B$-mode polarization measurements remain largely insensitive to SED variations \citep{prince_2024}, there may be a significant impact for upcoming experiments such as the Simons Observatory (SO) that will make use of deep, wide-area sky maps. 
This expectation is supported by recent findings from the {\scshape{Spider}} experiment, which identified anisotropic variations in the polarized dust SED across its 5\% sky patch \citep{spider_2025}.  \cite{vacher_2024} demonstrates that line-of-sight polarized dust SED variations larger than those predicted by benchmark simulations from the \cite{pan_exp_2025} are not excluded by the \emph{Planck} data.

The current best constraints on the primordial $B$-mode signal, derived from a combination of BICEP/Keck and \emph{Planck} data~\citep{bicepkeck_2021,tristram_2022}, are produced by a purely harmonic-domain power-spectrum-based analysis. Although sufficient for current instrumental noise levels and small ($\leq3\%$) sky coverage, power-spectrum-based Galactic foreground modeling may face challenges with future wide-area datasets \citep{remazeilles_2016, liu_2025}. 

Future analyses should therefore concentrate on map-based inference to make use of the anisotropic properties of the foreground emission. Broadly speaking, map-based methods may be divided into fully parametric approaches \citep{brandt_1994, eriksen_2006, stompor_2009} that explicitly model all sky components at the field level and internal linear combination (ILC) methods that are partly non-parametric \citep{bennett_2003}. 

Parametric ``field-level'' approaches, see e.g.\  \cite{beyond_planck_2023} for a recent example, explicitly model both foregrounds and instrumental effects and map out the entire field-level posterior using Markov chain Monte Carlo (MCMC) sampling. This explicit Bayesian approach
provides consistent propagation of uncertainties, a highly interpretable estimate, and statistical optimality in the case of a correct data model. However, all aspects of the data model have to be modeled accurately;  mismodeling will lead to biases. In practice, this is difficult. For example, the method requires an accurate statistical description of the noise, which has not yet been shown to be possible for modern ground-based experiments. In addition, the known non-Gaussian spatial distribution of the foreground emission \citep{miville_2007,regaldo_2020} is not captured by the Gaussian priors on the foreground emission that the method has to impose due to numerical constraints. Finally, these likelihood-based methods have to sample the entire latent space, e.g.\ the value of each sky component in each pixel, in order to obtain the marginal posterior distribution of the cosmological parameters of interest. The necessary repeated evaluation of the field-level likelihood (or its gradients) by the MCMC-based method is an enormous computational challenge that is inherently sequential, which prevents speedups using massively parallel architectures.

The complementary, non-parametric ILC method finds a minimum-variance point estimate of the CMB sky using an empirical covariance matrix that describes the total covariance (signal and noise) between all available frequency bands. 
Although less optimal compared to parametric methods due to the lack of prior information on the SEDs and spatial structure of the foregrounds, the ILC method is potentially more robust because it makes fewer assumptions about the contaminants. 
A notable variant is the Needlet-ILC (NILC) algorithm \citep{delabrouille_2009} that estimates the ILC covariance matrix in a wavelet basis in order to capture how the spatial correlation structure of the noise and signal varies over the sky. In practice, however, ILC-based inference pipelines, including those using the NILC method, produce biased constraints on $r$ when applied to foregrounds with significant SED anisotropy (see \cite{wolz_2024} as a recent example), as residual foreground contamination is not quantified by the point estimate.  The constrained-ILC \citep{chen_2009,remazeilles_2011, remazeilles_2021} method is a semi-blind approach that partly addresses this issue by jointly estimating the CMB and assumed foreground components and then marginalizing over the latter. This reduces bias if the assumed foreground SEDs are accurate, but does not address the fundamental limitation that the point estimate cannot quantify the residual contamination. Furthermore, the marginalized CMB estimate generally has significantly increased variance compared to the regular (unconstrained) ILC estimate. The number of foreground components included in the constrained estimate must therefore be tuned to balance bias reduction against variance increase. See \cite{abylkairov_2021, carones_2024} for approaches to automate this trade-off.

The three SO analysis pipelines described in \cite{wolz_2024} provide unbiased $r$ results for foreground models with modest amounts of anisotropic SED variations. When applied to the more challenging \texttt{d10s5} medium-complexity simulations from \cite{pan_exp_2025}, approximately unbiased results can still be obtained with the goal/optimistic SO SAT noise levels, but only when extending the nominal power-spectrum pipeline to the $C_{\ell}$-moments method from \cite{azzoni_2021}, reducing the  bias to just $1.0\sigma$, and adding approximate marginalization over dust contamination to the map-based parametric pipeline, achieving a $0.3 \sigma$ bias. These extensions increase $\sigma(r)$ by a factor 1.5 and 2.0, respectively. The NILC pipeline, for which no extension was considered, becomes biased at $8.3 \sigma$. 
\cite{liu_2025} further demonstrate  that the bias of the $C_{\ell}$-moments power-spectrum-based pipeline rapidly grows as the $\texttt{d10}$ SED anisotropy is artificially increased. 
No SO collaboration studies have presented results for the high-complexity $\texttt{d12}$ dust model, which, in addition to the strong spatial SED anisotropies of the $\texttt{d10}$ model, contains line-of-sight SED distortions \citep{martinez_2018}. However, \cite{bianchini_2025} reports biases ranging from 0 to $1\sigma$ for the power-spectrum-based method and 1 to $3\sigma$ for map-based parametric and ILC pipelines for  proposed CMB-S4 observations over a small, 3\%, sky patch for $\texttt{d12}$. Similarly, \cite{carones_2023} finds substantial foreground residuals around the degree-scale angular scales targeted by ground-based observatories for simulated LiteBIRD data based on the $\texttt{d12}$ model. While direct comparison across experiments is complicated by differences in frequency coverage, noise levels, and sky area, these results collectively suggest that current analysis methods are insufficient for wide-area ($f_{\mathrm{sky}}\geq 10\%$) ground-based observations in the presence of foregrounds with large, complex SED anisotropies, such as the $\texttt{d12}$ model.

We argue that, given the potential complexity of the foregrounds, the robustness that the ILC approach offers against unspecified contaminants is a valuable feature that should be exploited.  However, current applications of the ILC and constrained-ILC methods remain inherently limited in their usefulness because (i): they rely solely on the (marginalized) CMB estimate, discarding information in the data about the residual foreground contamination, and (ii), they rely on a point estimate with poorly understood statistical properties, which limits meaningful error propagation. 

Building upon \cite{surrao_2024a,surrao_2024}, we construct a simulation-based inference (SBI) framework for estimating $r$ that combines the robustness of the ILC method with the formal Bayesian treatment of parametric field-level inference.\footnote{All code used to produce the results of this paper is available at \url{https://github.com/AdriJD/sbi_bmode}.} We embed the NILC method into a Bayesian inference pipeline that avoids ad hoc assumptions about the likelihood or residual foreground contamination. The SBI framework allows joint use of the NILC estimates of the CMB, dust, and synchrotron component maps, including their spatial SED variations, and circumvents the difficulty one would face when constructing an accurate analytical likelihood of these correlated estimates, which are all derived from the same data.  In this sense, our approach extends the hybrid methods of \cite{azzoni_2023} and \cite{Carones_2025} to a fully Bayesian formulation. We explore whether unbiased $r$ inference can be achieved using relatively simple, cheap to generate,  simulations by employing data compression designed to be robust against mismodeling of Galactic foregrounds and instrumental noise. This exploits the ability of SBI to perform inference with data that are compressed in a complex manner.

A secondary goal of this exploratory work is to investigate the feasibility of a complete simulation-based $r$ inference pipeline for a ground-based experiment like SO. If the presented SBI setup allows for unbiased results in the presence of realistic foregrounds, it will provide a natural framework to marginalize over realistic instrumental systematics as well as include complex models of foreground emission that cannot be described by simple closed-form expressions. One example would be the incorporation of external tracers of the 3D distribution of Galactic dust, similar to \cite{martinez_2018}. 
Such models have so far only been used as validation data for traditional likelihood-based analyses, which cannot easily be adapted to use them directly  during inference. SBI, in contrast, can learn from simulations drawn from the models, allowing their complexities to be integrated into the inference process.

\section{Setup}

\subsection{Experimental setup}\label{sec:exp_setup}

For this exploratory paper, we adopt a relatively simple foreground model and experimental setup. We assume that the uncompressed data $\bm{d}$ are in the form of a pair of Stokes $Q$ and $U$ maps for each of the frequency bands of a generic ground-based observatory. The experiment roughly follows the SO Small Aperture Telescopes (SAT) specifications from \cite{so_forecast_2019} and observes in delta-function bands centered on $27$, $39$, $93$, $145$, $225$, and \SI{280}{\giga\hertz}. We augment the dataset with bands at $23$ and \SI{353}{\giga\hertz} that mimic the \emph{WMAP} K and \emph{Planck} 353 GHz channels, respectively. See Appendix~\ref{app:instrument} for more details. 

The data are related to the sky signal and noise as follows:
\begin{align}\label{eq:cmb_data_model}
    \bm{d} = \mathbf{M}\left(\widetilde{\mathbf{Y}} \mathbf{F} \mathbf{B} \bm{s} + \bm{n}\right) \, . 
\end{align}
Here, $\bm{s}$ is a set of $B$-mode spherical harmonic coefficients that represent both the CMB and the dust and synchrotron foregrounds.  The noise $\bm{n}$ is a set of statistically isotropic Stokes $Q$ and $U$ maps drawn from a power spectrum that approximates the noise of the SO SAT, \emph{WMAP} K, and \emph{Planck} 353~GHz bands. The  beam operator $\mathbf{B}$ describes the convolution with an azimuthally symmetric Gaussian beam that varies per frequency band. The matrix $\mathbf{F}$ describes a convolution with an  azimuthally symmetric high-pass filter that approximates the time-domain filtering operations typically applied by ground-based CMB observatories. The matrix $\widetilde{\mathbf{Y}}$ denotes the spin-weighted spherical harmonic transformation from the $B$-mode coefficients to Stokes $Q$ and $U$ maps \citep{zaldarriaga_1997,kamionkowski_1997}, see Appendix~\ref{app:sht}.  The matrix $\mathbf{M}$ denotes an optional sky mask. See Appendix~\ref{app:instrument} for further details on the instrumental setup. 

The $B$-mode signal $\bm{s}_{\nu}$, with frequency index $\nu$ written explicitly for clarity, is represented by the following set of spherical harmonic coefficients:
\begin{align}\label{eq:dust_total}
    \bm{s}_{\nu} =  \sum_{c} \mathbf{Y}^{\dagger}\mathbf{W} \mathbf{Q}_{\nu, c} \mathbf{Y} \bm{a}_{c} \, ,
\end{align}
where the sum runs over the sky components: $c \in \{\mathrm{CMB}, \mathrm{dust}, \mathrm{synchrotron}\}$. The $\bm{a}_{c}$ are spherical harmonic coefficients describing the amplitude field for each sky component. The matrix $\mathbf{W}$ is a diagonal matrix of integration weights for each pixel and $\mathbf{Q}$ is an $N_{\nu} \times N_{c}$ block matrix of $N_{\mathrm{pix}} \times N_{\mathrm{pix}}$ matrices that specifies the contribution of a given sky component $c$ to a given frequency band~$\nu$. The $\mathbf{Y}$ matrix denotes a spin-0 spherical harmonic transform, see Appendix~\ref{app:sht}.

 We assume that the data are calibrated with respect to the CMB blackbody spectrum and expressed in CMB-referenced thermodynamic units. Therefore, $\mathbf{Q}_{\nu, \mathrm{CMB}} = \mathbf{1} \, \forall \nu$ and the CMB contribution to $\bm{s}$ is simply given by $\bm{a}_{\mathrm{CMB}}$. These amplitudes are drawn from a $B$-mode power spectrum that is modeled as:
\begin{align}\label{eq:cmb_cl}
    C_{\ell} = rC^{h}_{\ell} + A_{\mathrm{lens}}C^{\zeta}_{\ell} \, ,
\end{align}
where $C^{h}_{\ell}$ denotes the primordial tensor contribution to the $B$-mode power spectrum and $C^{\zeta}_{\ell}$ denotes the $B$-mode power spectrum of lensed scalar modes. 

The components of $\mathbf{Q}$ that describe the dust and synchrotron foregrounds are more complex due to the spatial variations in the foreground SEDs. 
The parametrization of the foregrounds  closely follows that of \cite{azzoni_2021}. Starting with the dust component, the column of the $\mathbf{Q}$ matrix that governs the dust is given by:
\begin{align}
\left(\mathbf{Q}\right)_{\nu, \mathrm{dust}} =   \frac{\bm{\mu}(\nu, \bm{\beta}_{\mathrm{d}}, T_{\mathrm{d}})}{\bm{\mu}(\nu_{0,\mathrm{d}}, \bm{\beta}_{\mathrm{d}}, T_{\mathrm{d}})} \, .
\end{align}
The frequency-dependent factor $\bm{\mu}$ is given by the following modified blackbody spectrum:
\begin{align}\label{eq:dust_sed}
    \bm{\mu}(\nu, \bm{\beta}_{\mathrm{d}}, T_{\mathrm{d}}) =  \nu^{\bm{\beta}_{\mathrm{d}} - 2} B_{\nu}(T_{\mathrm{d}}) f(\nu) \, ,
\end{align}
where $B_{\nu}$ is the Planck function and $f(\nu)$ is a unit conversion factor converting from Rayleigh-Jeans to CMB-referenced temperature:
\begin{align}
f(\nu) = \frac{(\mathrm{e}^x - 1)^2}{x^2 \mathrm{e}^x} \, ,
\end{align}
with $x = h \nu / (k_\mathrm{B} T_{\mathrm{CMB}})$. The dust reference frequency $\nu_{0,\mathrm{d}}$ is set to $\SI{353}{\giga\hertz}$. 
Since the frequency bands we consider lie in the Rayleigh-Jeans tail of the dust spectrum, variations in the dust temperature $T_{\mathrm{d}}$ do not, to first-order approximation, change the shape of the spectrum and thus have a minor effect on the inference on $r$ \citep{liu_2025}. We therefore fix $T_{\mathrm{d}} = \SI{19.6}{\kelvin}$ throughout this work.
The $\bm{\beta}_{\mathrm{d}}$ vector of spectral indices is given by $\mathbf{Y} \bm{b}$, where the spherical harmonic coefficients $\bm{b}$ are drawn from the following angular power spectrum:
\begin{align}\label{eq:gamma_dust}
    C^{\beta_{\mathrm{d}}}_{\ell} = B_{\mathrm{d}} \left(\frac{\ell}{\ell_0}\right)^{\gamma_\mathrm{d}} \, ,
\end{align}
where we use $\ell_0 = 1$. We set the monopole of the $\bm{\beta}_{\mathrm{d}}$ map  equal to the sky-averaged index parameter~$\beta_{\mathrm{d}}$. The dust amplitudes $\bm{a}_{\mathrm{d}}$ in Eq.~\eqref{eq:dust_total} are drawn from a power-law power spectrum $C^{\mathrm{d}}_{\ell}$ with power-law index $\alpha_d$, defined as follows:
\begin{align}\label{eq:dust_ell_shape}
    \frac{\ell (\ell + 1)C^{\mathrm{d}}_{\ell}}{2 \pi} = A_{\mathrm{d}}  \left(\frac{\ell}{80}\right)^{\alpha_d} \, .
\end{align}

The synchrotron column of the $\mathbf{Q}$ matrix is given by:
\begin{align}
\left(\mathbf{Q}\right)_{\nu, \mathrm{sync.}} =   \frac{\bm{\omega}(\nu, \bm{\beta}_{\mathrm{s}})}{\bm{\omega}(\nu_{0,\mathrm{s}}, \bm{\beta}_{\mathrm{s}})} \, ,
\end{align}
with SED factor:
\begin{align}\label{eq:sync_sed}
    \bm{\omega}(\nu, \bm{\beta}_{\mathrm{s}}) =  \nu^{\bm{\beta}_{\mathrm{s}}} f(\nu) \, .
\end{align}
The synchrotron reference frequency $\nu_{0,\mathrm{s}}$ is $\SI{23}{\giga \hertz}$. The spatial correlation structure of the synchrotron spectral index $\bm{\beta}_{\mathrm{s}}$ is modeled similarly to Eq.~\eqref{eq:gamma_dust}:
\begin{align}\label{eq:gamma_sync}
    C^{\beta_{\mathrm{s}}}_{\ell} = B_{\mathrm{s}} \left(\frac{\ell}{\ell_0}\right)^{\gamma_{\mathrm{s}}} \, ,
\end{align}
with $\ell_0 = 1$. The average of the $\bm{\beta}_{\mathrm{s}}$ map is set to the sky-averaged index parameter~$\beta_{\mathrm{s}}$. 
The spatial correlation structure of the synchrotron amplitudes $\bm{a}_{\mathrm{s}}$ has the same functional form as Eq.~\eqref{eq:dust_ell_shape}, but uses $A_{\mathrm{s}}$ and $\alpha_{\mathrm{s}}$ instead:
\begin{align}\label{eq:sync_ell_shape}
    \frac{\ell (\ell + 1)C^{\mathrm{s}}_{\ell}}{2 \pi} = A_{\mathrm{s}}  \left(\frac{\ell}{80}\right)^{\alpha_s} \, .
\end{align}

Finally, in order to incorporate the known spatial correlation between the polarized dust and synchrotron emission \citep{choi_2015}, we include a correlation between the dust and synchrotron amplitudes ($\bm{a}_{\mathrm{d}}$ and $\bm{a}_{\mathrm{s}}$) described by the following cross-power spectrum:
\begin{align}\label{eq:rho_ds}
    C^{\mathrm{ds}}_{\ell} = \rho_{\mathrm{ds}}\sqrt{C_{\ell}^{\mathrm{d}} C_{\ell}^{\mathrm{s}}} \, ,
\end{align}
with a scalar correlation coefficient $\rho_{\mathrm{ds}} \in [-1,1]$.

\subsection{Neural posterior estimation}

We start by sketching the statistical inference problem. Take $\bm{\theta}$ to be a set of hyperparameters that govern the cosmological and foreground models, e.g.\ $r$, $A_{\mathrm{lens}}$, $A_{\mathrm{d}}$, $\gamma_{\mathrm{d}}$, etc., and $\bm{d}$ to be data, e.g.\ the sky maps at each experimental frequency band. The data-generating process, i.e.\ the likelihood  $P(\bm{d}|\bm{\phi})$, depends on a second set of parameters $\bm{\phi}$ that, in turn, depend on $\bm{\theta}$. In our case, the $\bm{\phi}$ parameters should be understood as e.g.\ the per-pixel CMB and foreground amplitudes for a particular realization of the sky. It is clear that the dimensionality of $\bm{\phi}$ is typically extremely large, $\mathcal{O}(10^{6})$ in our case, similar in size to the data vector $\bm{d}$.

The hierarchical nature of the problem allows the posterior of the parameters of interest, $\bm{\theta}$, to be expressed as follows:
\begin{align}
\begin{split}\label{eq:hierarchical_model}
P(\bm{\theta} | \bm{d}) &= \int  P(\bm{\theta}, \bm{\phi} | \bm{d}) \mathrm{d}\bm{\phi} \, , \\
&\propto \int P(\bm{d} | \bm{\phi}) P(\bm{\phi}| \bm{\theta}) P(\bm{\theta}) \mathrm{d}\bm{\phi} \, . 
\end{split}
\end{align}
If the densities in Eq.~\eqref{eq:hierarchical_model} and their derivatives with respect to $\bm{\theta}$ and $\bm{\phi}$ can be evaluated, one can in principle use an MCMC method such as Hamiltonian Monte Carlo (HMC) \citep{duane_1987} to sample from the joint posterior $P(\bm{\theta}, \bm{\phi} | \bm{d})$. With the joint samples in hand, one may then estimate the integral over $\bm{\phi}$ to obtain an estimate of $P(\bm{\theta} | \bm{d})$, the marginalized posterior distribution we are ultimately interested in. 
It should be noted that this is challenging: the MCMC chain has to fully explore the high-dimensional parameter space spanned by the $\bm{\phi}$ parameters that contribute to the integral. 
In fact, an HMC-based sampling of the joint CMB and foreground posterior has not yet been shown to be feasible. In practice, only MCMC sampling in the form of Gibbs sampling, which requires that closed-form sampling algorithm exists for the different conditional distributions of $P(\bm{\theta}, \bm{\phi} | \bm{d})$, has been successfully applied, e.g.\ \cite{planck_2015_x_diffuse, beyond_planck_2023}. 

In this paper, we use simulation-based inference (SBI) to focus on the case where HMC or Gibbs sampling cannot be used, for example, the case in which the generative model for the foregrounds is complicated enough that $P(\bm{\phi}|\bm{\theta})$ cannot be evaluated, only sampled from. We assume that we can still produce simulated versions of the data:
\begin{align}\label{eq:forward_model_gen}
\begin{split}
    \bm{\theta}_i &\sim P(\bm{\theta}) \, ,\\
    \bm{\phi}_i &\sim P(\bm{\phi} | \bm{\theta}_i) \, ,\\
    \bm{d}_i &\sim P(\bm{d}| \bm{\phi}_i) \, ,
\end{split}
\end{align}
where we use the $\sim$ symbol to denote the act of sampling from a distribution.
SBI uses the simulated pairs of parameters and data vectors $\{\bm{\theta}_i, \bm{d}_i \}_{i=1}^{N_{\mathrm{sim}}}$ in combination with the actual observed data to directly estimate the marginal posterior $P(\bm{\theta}|\bm{d})$. In doing so, it avoids the need to evaluate the $P(\bm{d} | \bm{\phi})$ and $P(\bm{\phi} | \bm{\theta})$ PDFs in Eq.~\eqref{eq:hierarchical_model} and bypasses the explicit marginalization over the high-dimensional latent CMB and foreground fields included in~$\bm{\phi}$.

The type of SBI that we use to estimate the marginal posterior of $\bm{\theta}$ is referred to as neural posterior estimation (NPE) \citep{papamakarios_2016}. In particular, we adopt NPE based on conditional normalizing flows. This type of NPE is inherently limited to inference problems with relatively low-dimensional parameter and data spaces. The restriction on parameter dimensionality is not a major limitation as we are interested in the $\mathcal{O}(10)$ parameters that make up $\bm{\theta}$. Similarly, the constraint of a small data vector is not problematic because SBI remains well-defined regardless of the choice of compression of the data.\footnote{With a converged and sufficiently expressive posterior estimator, SBI yields the correct posterior for data consistent with the assumed forward model regardless of the choice of compression.}

This property, the fact that SBI is well-defined  regardless of the choice of compression, is a  central aspect our work. Since the data compression $\bm{x} = f(\bm{d})$, with $f : \mathbb{R}^{N_{\bm{d}}} \rightarrow \mathbb{R}^{N_{\bm{x}}}$ for $N_{\bm{x}} \ll N_{\bm{d}}$, can be freely specified, we construct a compression scheme that reduces sensitivity to aspects of the data for which the model is less reliable, e.g.\ the detailed noise properties, the exact functional forms of the foreground SEDs and the anisotropic and non-Gaussian properties of the spatial distribution of the foregrounds. To achieve this, we pick a nonlinear compression scheme consisting of the NILC algorithm followed by a cross-power spectrum estimate and a binning operation. This procedure introduces no additional free parameters. See Sec.~\ref{sec:nilc} below for  details on the compression scheme. 

The optimization of $\bm{\lambda}$, the parameters of the normalizing flow $q_{\bm{\lambda}}$ used for the NPE, is done by minimizing the forward Kullback-Leibler divergence, $D_{\mathrm{KL}}$, between the joint distribution of parameters and compressed data, $P(\bm{\theta}, \bm{x})$, and our approximation of that same distribution: $q_{\bm{\lambda}}\bigl(\bm{\theta} | \bm{x}\bigr)P(\bm{x})$ \citep{papamakarios_2019}:
\begin{align}\label{eq:kl_divergence}
\begin{split}
    \argmin_{\bm{\lambda}} \, &D_{\mathrm{KL}}\biggl( P(\bm{\theta} , \bm{x}) \, \, || \, \, q_{\bm{\lambda}}\bigl(\theta | \bm{x}\bigr) P(\bm{x}) \biggr) \\
    &= \argmin_{\bm{\lambda}} \int  P(\bm{\theta}, \bm{x}) \log \frac{P(\bm{\theta}| \bm{x})}{q_{\bm{\lambda}}\bigl(\bm{\theta} | \bm{x}\bigr)} \mathrm{d}\bm{\theta}  \mathrm{d}\bm{x} \, , \\
    &= \argmin_{\bm{\lambda}} \int  -P(\bm{\theta}, \bm{x}) \log q_{\bm{\lambda}}\bigl(\bm{\theta} | \bm{x}\bigr) \mathrm{d}\bm{\theta}  \mathrm{d}\bm{x} \, , \\
    &\approx \argmin_{\bm{\lambda}} \frac{1}{N_{\mathrm{sim}}}\sum_{i=1}^{N_{\mathrm{sim}}} - \log q_{\bm{\lambda}}\bigl(\bm{\theta}_i | \bm{x}_i\bigr) \, .
\end{split}
\end{align}
The sum introduced in the final step runs over draws $\bm{\theta}_i, \bm{x}_i$ from the joint distribution $P(\bm{\theta}, \bm{x})$ that are provided by the forward model from Eq.~\eqref{eq:forward_model_gen} followed by the compression step $\bm{x}_i = f(\bm{d}_i)$. In the limit of large $N_{\mathrm{sim}}$ the approximation should converge to an equality. Crucially, the optimization does not require the evaluation of $P(\bm{\theta}, \bm{x})$, $P(\bm{\theta}| \bm{x})$, $P(\bm{x} | \bm{\theta})$, or $P(\bm{x})$.

The architecture of the normalizing flow we use for $q$ is a masked autoregressive flow (MAF) \citep{papamakarios_2017}. Normalizing flows are well-suited to NPE since they enable both efficient density evaluation, required for the minimization in Eq.~\eqref{eq:kl_divergence}, and sampling, which is required to obtain samples of the posterior. 
They model a probability distribution using a composition of invertible and differentiable transformations that map a standard Gaussian base density to the desired non-Gaussian distribution. Sampling is  performed by drawing from the base distribution and applying the transformation, while density evaluation is done by evaluating the Gaussian distribution in combination with a change of variables computed using the inverse transform and Jacobian of the inverse transform. See \cite{papamakarios_2019} for a pedagogical introduction and an overview of the different flow architectures. 

As long as the normalizing flow $q$ and $N_{\mathrm{sim}}$ are expressive and large enough, the minimization in Eq.~\eqref{eq:kl_divergence} will result in a density estimate that converges to the  $P(\bm{\theta} | \bm{x})$ posterior. We check the convergence of the normalizing flow using coverage tests in Sec.~\ref{sec:validation}. Once trained, we take the actual compressed observed data $\bm{x}_{\mathrm{obs}} = f(\bm{d}_{\mathrm{obs}})$ and sample from $q_{\bm{\lambda}}(\bm{\theta} | \bm{x}_{\mathrm{obs}})$ to obtain the posterior samples. 
The minimization in Eq.~\eqref{eq:kl_divergence} is performed using the Adam optimizer \citep{kingma_2014}. We use the Optuna optimization library \citep{akiba_2019} to tune the hyperparameters of Adam and the MAF architecture, see Appendix~\ref{app:hyperparam_tuning} for details. 
 
\subsection{Needlet-ILC as a compression algorithm}\label{sec:nilc}

The overarching idea of our compression strategy is to not rely on NILC as an unbiased estimator of the CMB, as is mostly done in the literature, but simply rely on it as a practical and effective compression algorithm. Specifically, NILC reduces the $N_{\nu}$ frequency maps into a smaller set of more informative sky component maps. By operating in a wavelet basis, the method suppresses the impact of the statistical anisotropy of Galactic foregrounds, thereby reducing biases arising from mismodeling between simulations and data. In contrast to ``black-box'' neural compression schemes commonly used in SBI, NILC has a robust and physically motivated optimization criterion: minimizing the variance of any signal that does not follow the prescribed SEDs. 
This simple criterion increases robustness against mismodeling between simulations and data, and avoids introducing additional learnable free parameters whose training would risk undermining that robustness.

Given an $N_{\nu} \times 2\times N_{\mathrm{pix}}$ Stokes $Q$ and $U$ data vector $\bm{d}$, the ILC $B$-mode estimate of a set of sky components is given by:
\begin{align}\label{eq:ilc}
    \hat{\bm{s}} = \mathbf{E} \left(\mathbf{A}^{\top} \mathbf{C}^{-1} \mathbf{A}\right)^{-1} \mathbf{A}^{\top} \mathbf{C}^{-1} \widetilde{\bm{d}} \, ,
\end{align}
where $\widetilde{\bm{d}} = \mathbf{G}\mathbf{Y}\widetilde{\mathbf{Y}}^{\dagger}\mathbf{W} \bm{d}$ are the data transformed to $E$- and $B$-mode maps, followed by $\mathbf{G} = \mathrm{diag}(0, 1)$, which selects the $B$-mode component. The output $\hat{\bm{s}}$ is an $N_c \times N_{\mathrm{pix}}$ vector of estimated $B$-mode sky components, indexed by $c$. The matrix $\mathbf{A}$ is an $N_{c} \times N_{\nu}$ block matrix, where each block is an $N_{\mathrm{pix}} \times N_{\mathrm{pix}}$ matrix. The rows, indexed by $c$, contain the assumed response of each frequency band to sky component $c$. In our simple setup, this response is given by a single scalar value: the SED of the sky component evaluated at the frequency of the band; we assume no spatial variations or couplings between pixels.\footnote{An actual analysis should integrate the SED of the sky component over the frequency passbands of the instrument and take into account the frequency dependence of the beam \citep{madhavacheril_2020}.} Different choices for the shape of the $\mathbf{A}$ matrix have received different designations in the literature. 
When $N_{c}=1$, i.e.\ only a single sky component is inferred, the algorithm is referred to as the ILC algorithm~\citep{bennett_2003}. The constrained ILC algorithm \citep{chen_2009,remazeilles_2011, remazeilles_2021} has $N_{c}>1$, which means that $N_{c}-1$ components are ``deprojected'' from the estimate. The deprojection is quantified by the matrix $\mathbf{E}$, which is an $N_{c} \times N_{c}$ diagonal block matrix that selects the sky component of interest.\footnote{In the case where $\widetilde{\bm{d}} \sim \mathcal{N}(\mathbf{A}\bm{s}, \mathbf{C})$ with known $\mathbf{C}$, $\hat{\bm{s}}$ in Eq.~\eqref{eq:ilc} is a Gaussian field with mean $\mathbf{E}\bm{s}$ and covariance $\mathbf{E}\left(\mathbf{A}^{\top} \mathbf{C}^{-1}\mathbf{A}\right)^{-1}\mathbf{E}^{\top}$. Deprojecting can therefore be understood as analytically marginalizing (removing entries from the covariance matrix and mean vector) over the unwanted sky components under the assumption of Gaussian statistics.} For our proposed method, which follows the NILC SBI approach introduced in \cite{surrao_2024}, 
$N_c > 1$, but we set $\mathbf{E}$ to the identity matrix, keeping all $N_{c}$ components specified by $\mathbf{A}$. For clarity, we will refer to this method as ``Joint ILC''. 

The matrix $\mathbf{C}$ in Eq.~\eqref{eq:ilc} denotes the total covariance of the data, i.e. the covariance of the CMB, any foregrounds with specified SEDs, unspecified foregrounds, and instrumental noise. In order to include the contribution from unknown sky components, it is important that the matrix is estimated from the data itself. In this sense the method is ``blind'': it will minimize variance due to a priori unspecified contaminants.\footnote{This is in contrast to parametric methods that use the inverse noise covariance $\mathbf{N}^{-1}$ instead of inverse total covariance $\mathbf{C}^{-1}$.} 
ILC methods are thus non-linear in the input data: they apply an empirical inverse covariance matrix to the same data it was estimated from. When ignored, this non-linearity causes a noticeable bias, the so-called ILC-bias \citep{delabrouille_2009}. For traditional methods that rely on the ILC maps as unbiased estimates of the sky, this bias has to be mitigated using a bias-variance trade-off method \citep{delabrouille_2009,mccarthy_2024,coulton_2024} and any remaining bias has to be estimated using simulations \citep{basak_2012}. In our SBI setup, no post hoc corrections are necessary: any remaining bias after bias mitigation is learned from the simulations in a consistent and automated way.

The ``Needlet'' in Needlet-ILC refers to the wavelet basis that is used for the $\mathbf{C}$ matrix in Eq.~\eqref{eq:ilc}. Given that wavelets are localized in both harmonic and spatial domains, NILC is tailored towards foregrounds and noise that vary both as a function of angular scale and sky position.  
Specifically, the weighting that is applied to the data, see Eq.~\eqref{eq:ilc}, is allowed to vary over the sky to adapt to the anisotropy of the foreground and/or noise. As a result, the CMB estimate should be less contaminated by residual foregrounds and more robust to foreground mismodeling in the simulations. 
For comparison, we also use Harmonic-ILC (HILC), which uses a purely harmonic basis for the $\mathbf{C}$ matrix, for which this argument should not hold.  HILC can capture scale dependence of foreground contaminants, but NILC can capture both scale dependence and pixel dependence. For the NILC method, the covariance matrix is given by:
\begin{align}\label{eq:nilc_cov}
\mathbf{C} = \sum_{j=1}^{N_{\mathrm{wav}}} \sum_{j'=1}^{N_{\mathrm{wav}}}\mathbf{Y}\mathbf{K}_{j}^{\dagger}  \mathbf{Y}^{\dagger} \mathbf{W} \mathbf{N}_{jj'} \mathbf{W} \mathbf{Y} \mathbf{K}_{j'} \mathbf{Y}^{\dagger}  \, ,
\end{align}
where the $\mathbf{K}$ matrices are $N_{\mathrm{wav}}\times1$ block matrices that contain the wavelet kernels. These are diagonal $N_{\mathrm{harm}} \times N_{\mathrm{harm}}$ matrices with the harmonic response of the wavelet on the diagonal. The wavelet kernels are constrained such that $\sum_j \mathbf{K}_{j}^{\dagger}\mathbf{K}_{j} = \mathbf{1} \ \forall j$. The $\mathbf{N}_{jj'}$ covariance matrix is estimated from the data as specified in \cite{mccarthy_2024}.\footnote{In our notation, the empirical covariance matrix is given by $\mathbf{N}_{jj'} = \mathbf{Y}\mathbf{K}_{j} \mathbf{Y}^{\dagger} \mathbf{W} \widetilde{\bm{d}} \widetilde{\bm{d}}^{\dagger} \mathbf{W} \mathbf{Y} \mathbf{K}_{j'}^{\dagger} \mathbf{Y}^{\dagger}$, where each element indexed by $(j, j')$ is an $(N_{\nu} \times N_{\mathrm{pix}}) \times (N_{\mathrm{pix}} \times N_{\nu})$ matrix. The $j\neq j'$ and off-diagonal pixel elements are set to zero and the diagonal of the pixel dimensions is smoothed to reduce the ILC bias. See \cite{mccarthy_2024} for details.} 
The covariance matrix for the HILC method is given by:
\begin{align}
    \mathbf{C} = \mathbf{Y} \mathbf{N}^{\mathrm{H}} \mathbf{Y}^{\dagger} \, ,
\end{align}
where $\mathbf{N}^{\mathrm{H}}$ is an $(N_{\nu}\times N_{\mathrm{harm}}) \times (N_{\mathrm{harm}} \times N_{\nu})$  matrix.\footnote{The matrix is estimated from the data as $\mathbf{N}^{\mathrm{H}} = \mathbf{Y}^{\dagger} \mathbf{W} \widetilde{\bm{d}} \widetilde{\bm{d}}^{\dagger} \mathbf{W} \mathbf{Y}$. The elements with $\ell \neq \ell'$ and $m \neq m'$ are neglected and the diagonal is smoothed to reduce the ILC bias \citep{mccarthy_2024}.}

An SED has to be provided for each of the $N_c$ sky components included in the $\hat{\bm{s}}$ estimate in Eq.~\eqref{eq:ilc}. For our main results, we include five components.  First, we include a component with the CMB blackbody SED. Second, we include a dust-like component with SED given by $\mu(\nu, \bar{\beta}_{\mathrm{d}}, \bar{T}_{\mathrm{d}})$, where $\mu$ is a scalar version of the SED given by Eq.~\eqref{eq:dust_sed}. Third, we include a synchrotron-like component with SED given by $\omega(\nu, \bar{\beta}_{\mathrm{s}})$, the scalar version of Eq.~\eqref{eq:sync_sed}. Here, $\bar{\beta}_{\mathrm{d}}$, $\bar{\beta}_{\mathrm{s}}$, and $\bar{T}_{\mathrm{d}}$ are fiducial values for the dust and synchrotron spectral indices and the dust temperature, respectively. For the last two components, we make use of the moment expansion method from \cite{chluba_2017}. In our case, this approach amounts to linearizing the SED of the foregrounds with respect to the dust and synchrotron spectral indices. The linear deviations from their fiducial values, $\bar{\beta}_{\mathrm{d}}$ and $\bar{\beta}_{\mathrm{s}}$, can be treated as two additional sky components in the $\mathbf{A}$ matrix in Eq.~\eqref{eq:ilc}:
\begin{align}\label{eq:delta_beta_dust}
(\mathbf{A})_{4,\nu} = \mathbf{1}\left.\frac{\partial}{\partial \beta_{\mathrm{d}}} \mu(\nu, \beta_{\mathrm{d}}, \bar{T}_{\mathrm{d}}) \right|_{\beta_{\mathrm{d}} = \bar{\beta}_{\mathrm{d}}} \, ,
\end{align}
and
\begin{align}
(\mathbf{A})_{5,\nu} = \mathbf{1}\left.\frac{\partial}{\partial \beta_{\mathrm{s}}} \omega(\nu, \beta_{\mathrm{s}}) \right|_{\beta_{\mathrm{s}} = \bar{\beta}_{\mathrm{s}}} \, .
\end{align}
We set the fiducial value for the dust spectral index parameter to $\bar{\beta}_{\mathrm{d}} = 1.5$, which is deliberately chosen to deviate from the true parameter values in our tests, the prior mean, and the best-fitting values for the PySM models, in order to emphasize that the method does not require a priori knowledge of the true value. 
For the synchrotron spectral index, we find that a fiducial value set to the prior mean, $\bar{\beta}_{\mathrm{s}} = -3$, provides significantly suboptimal inference results for the synchrotron parameters.\footnote{The effect of a suboptimal $\bar{\beta}_{\mathrm{s}}$ on the other parameters, including $r$ and $A_{\mathrm{lens}}$, is very minor due to the small contribution of the synchrotron signal to the data.} This occurs only when the $\mathbf{A}$ matrix includes the linear deviation of $\beta_{\mathrm{d}}$. Testing reveals that, given the relatively degenerate linear system that the ILC solves in this case, it is advantageous to pick a fiducial synchrotron SED that is more distinct from the other assumed SEDs. 
We therefore adopt  $\bar{\beta}_{\mathrm{s}} = -5$, which produces tighter constraints on the synchrotron parameters. For tests in Sec.~\ref{sec:results_joint_nilc} where we do not include the linear deviation of $\beta_{\mathrm{d}}$, we use the prior mean: $\bar{\beta}_{\mathrm{s}} = -3$. We only use a linear expansion in $\beta_{\mathrm{d}}$ and $\beta_{\mathrm{s}}$. We expect this to be sufficient for the SO noise levels, but the method generalizes straightforwardly to higher-order moments, or to the inclusion of dust moments with respect to $T_d$. This inclusion and the selection of optimal fiducial parameters would be interesting to explore in future work.

We further compress the data by computing the angular auto- and cross-spectra between the NILC sky component estimates:
\begin{align}\label{eq:cross_spectra}
    \hat{C}^{ij}_{\ell} = \frac{1}{2(2 \ell + 1)}\sum_{m = -\ell}^{\ell}  \left[ \hat{s}^i_{\ell m} (\hat{s}^j_{\ell m})^* + \hat{s}^j_{\ell m} (\hat{s}^i_{\ell m})^* \right]\, ,
\end{align}
where $i, j \in \{\mathrm{CMB, dust, synchrotron, \delta \beta_d, \delta\beta_s}\}$ and $\hat{s}^i_{\ell m}$ are given by the (spin-0) spherical harmonic transform of the $B$-mode sky component maps: $(\mathbf{Y}^{\dagger}\mathbf{W}\hat{\bm{s}} )^i_{\ell m}$. 
No attempt is made to undo the mode-coupling and bias in $\hat{C}^{ij}_{\ell}$ caused by the mask used in Sec.~\ref{sec:pysm_results}, as any bias will be learned from the simulations.\footnote{In our current setup, mode-coupling due to masking or other anisotropic weighting will cause some loss of optimality when left uncorrected. However, this is negligible for the mask geometry and bin width used in this paper.}
We only use cross-spectra between two versions of the data with independent noise contribution, as detailed further in Sec.~\ref{sec:sim_setup}. Finally, the cross-spectra are binned in $\ell$ as a last compression step. 

The compression into binned (cross-)power spectra serves two purposes. First, it achieves the bulk of the dimensionality reduction. Second, the inference becomes significantly less sensitive to noise modeling since we only use cross-correlations between observations with independent noise. This removes, on average, the noise contribution to Eq.~\eqref{eq:cross_spectra}. For ground-based experiments, where noise modeling is particularly  challenging, it is common practice for power-spectrum-based analyses to rely solely on cross-correlations between independent observations and exclude auto-correlations \citep{bennet_2003_ps}. Here, we adopt the same strategy.\footnote{In our case, where the data are mostly signal dominated, excluding the auto-correlations does not lead to a significant loss of information.} Thus, the compression can be understood as discarding the parts of the data that are overly sensitive to the quality of the noise modeling.  We emphasize that this is an advantage of SBI over field-level analyses, which necessarily have to model the noise accurately to avoid biases.\footnote{This advantage of SBI for CMB analyses was already pointed out by \cite{alsing_2019}.}

The inclusion of the foreground components in the $\mathbf{A}$ matrix in Eq.~\eqref{eq:ilc} suggests that the reconstructed CMB is free from foreground contamination \citep{remazeilles_2011}, making the  inclusion of the foreground components in the compressed data vector seem unnecessary. However, this intuition is incorrect. We argue, and demonstrate in Sec.~\ref{sec:results_joint_nilc}, that their inclusion serves two purposes. First, the foreground auto-spectra and their mutual cross-spectra contain information about the deviation from the fiducial model specified by $\overline{\beta}_{\mathrm{d}}$, $\overline{T}_{\mathrm{d}}$ and $\overline{\beta}_{\mathrm{s}}$, as well as the total power in the foreground realization, which affects the statistical uncertainty on~$r$. Second, the cross-spectra between the CMB component and the foreground components, which will generally not be zero due to misspecifications of the fiducial model assumed in $\mathbf{A}$, provide a measure of the amount of mismodeling and, crucially, a measure of the contamination in the CMB component, as highlighted and shown by~\cite{surrao_2024}.\footnote{A restricted version of this approach was explored in \cite{errard_2019, wolz_2024}. In the latter paper, it was found that the map-based parametric foreground cleaning method from \cite{poletti_2023,rizzieri_2025} requires marginalizing over scaled estimated dust power spectra to achieve unbiased $r$ estimates on the PySM $\texttt{d1}$ and $\texttt{d10}$ dust simulation with an instrumental setup similar to ours. \cite{Carones_2025} demonstrates that unbiased NILC likelihood-based inference with LiteBIRD applied to $\texttt{d1}$ and $\texttt{d10}$ is achieved by marginalizing over the amplitude of a residual dust power spectrum, a multi-frequency dust estimate that is weighted by the CMB NILC weights. While these approaches are conceptually similar to ours, they do not capture the correlations between the CMB and foreground estimates, which, as we will see, contain a substantial amount of useful information.} 

The NILC component maps should be viewed only as compressed representations rather than true sky component maps. As discussed above, the foreground components will be mixed and biased due to the misspecification of the foreground SEDs in the $\mathbf{A}$ matrix. Additionally, the weak correlation between the dust and synchrotron components \citep{choi_2015} introduces a small bias in the dust and synchrotron estimates that is proportional to the correlation coefficient \citep{delabrouille_2008,hurrier_2013}. Because our forward model includes the dust-synchrotron correlation, we  expect that this small bias will be automatically corrected for by the SBI. 
A similar effect may arise from correlations between the spatial variation of the spectral indices and the foreground amplitudes. 
In the current setup, we ignore this correlation in our simulations, as it appears to be negligible in the \emph{Planck} $B$-mode data \citep{liu_2025}. Future work should explore the effect of its inclusion in the forward model.

\subsection{Simulation details}\label{sec:sim_setup}

\begin{deluxetable}{c|lllc}
\tablecaption{Overview of priors. \label{table:prior}}
\tablehead{Param. & Distribution & PySM & Unit & Eq.
}
\startdata
     $r$ & $\mathcal{N}_{[0,\infty)}(0, 0.02)$ & - &  - & \eqref{eq:cmb_cl}\\
     $A_{\mathrm{lens}}$ & $\mathcal{N}_{[0,10]}(0.4, 0.2)$ & - & - & \eqref{eq:cmb_cl}\\
     $A_{\mathrm{d}}$ & $\mathcal{N}_{[12,44]}(28, 4)$ & $\mathcal{N}_{[12,88]}(28, 24)$ & $\SI{}{\micro\kelvin\squared}$ & \eqref{eq:dust_ell_shape}\\
    $\alpha_{\mathrm{d}}$ & $\mathcal{N}(-0.2, 0.2)$ & $\mathcal{N}(-0.2, 0.6)$ & - & \eqref{eq:dust_ell_shape}\\
    $\beta_{\mathrm{d}}$ & $\mathcal{N}(1.59, 0.2)$ & $\mathcal{N}(1.59, 0.4)$ & - & \eqref{eq:gamma_dust}\\
    $B_{\mathrm{d}}$ & $\mathcal{N}_{[0, 1.59]}(0.3, 0.5)$ & - & - & \eqref{eq:gamma_dust}\\
    $\gamma_{\mathrm{d}}$ & $\mathcal{N}_{[-6, -2]}(-4, 1)$ & - & - & \eqref{eq:gamma_dust}\\
     $A_{\mathrm{s}}$ & $\mathcal{N}_{[0,20]}(1, 1.5)$ & $\mathcal{N}_{[0,20]}(1, 3)$ & $\SI{}{\micro\kelvin\squared}$ &\eqref{eq:sync_ell_shape}\\
    $\alpha_{\mathrm{s}}$ & $\mathcal{N}(-0.7, 0.2)$ & $\mathcal{N}(-0.7, 0.4)$ & - &\eqref{eq:sync_ell_shape}\\
    $\beta_{\mathrm{s}}$ & $\mathcal{N}(-3, 0.2)$ & $\mathcal{N}(-3, 0.4)$ & - &\eqref{eq:gamma_sync}\\
    $B_{\mathrm{s}}$ & $\mathcal{N}_{[0, 0.5]}(0.3, 0.5)$ & - & - & \eqref{eq:gamma_sync}\\
    $\gamma_{\mathrm{s}}$ & $\mathcal{N}_{[-6, -2]}(-4, 1)$ & - & - & \eqref{eq:gamma_sync}\\
    $\rho_{\mathrm{ds}}$ & $\mathcal{N}(0, 0.2)$ & - & - & \eqref{eq:rho_ds}
\enddata
\tablecomments{We use $\mathcal{N}_{[0,\infty)}$ to denote a half-normal distribution and $\mathcal{N}_{[a, b]}$ for a truncated normal distribution defined in the interval $[a, b]$. The ``PySM'' column reports widened priors that are used for the application on the PySM foreground maps in Sec.~\ref{sec:pysm_results}. We only tabulate the priors that are updated.}
\end{deluxetable}

\begin{deluxetable}{l|l}
\tablecaption{Overview of compression setups \label{table:nilc_types}}
\tablehead{Name & \colhead{Maps/deprojected comps.} 
}
\startdata
    Joint NILC (\texttt{c+d+s+dbd+dbs})&CMB, dust, sync., $\delta \beta_{\mathrm{d}}$,  $\delta \beta_{\mathrm{s}}$ \\
    Joint NILC (\texttt{c+d+s})&CMB, dust, sync. \\
    NILC (\texttt{c}) & - \\
    Cnstr.\ NILC (\texttt{c+d+s+dbd+dbs})&Dust, sync., $\delta \beta_{\mathrm{d}}$, $\delta \beta_{\mathrm{s}}$\\
    Cnstr.\ NILC (\texttt{c+d+s})&Dust, sync.
\enddata
\tablecomments{For the joint NILC type, the rightmost column indicates into which sky component maps the frequency maps are compressed. For the constrained NILC type, where the data are always compressed into just a single CMB map, the column indicates the components that are deprojected from the CMB map. The joint NILC ($\texttt{c+d+s+dbd+dbs}$) setup is the default used throughout the paper.}
\end{deluxetable}

As described in the previous sections, generating the parameter-data pairs, $\{ \bm{\theta}_i, \bm{x}_i,\}_{i=1}^{N_{\mathrm{sim}}}$, involves four steps: (i) draw a parameter vector from the prior; (ii) generate sky component maps; (iii) simulate the data; and (iv) apply the joint NILC algorithm and compute the binned auto- and cross-spectra of the component maps. In this section, we provide implementation details for each step.

The prior is described in Table~\ref{table:prior}. To mimic a realistic analysis, we adopt wide, uninformative priors for all parameters except $B_{\mathrm{d}}$, $\gamma_{\mathrm{d}}$, $B_{\mathrm{s}}$, and $\gamma_{\mathrm{s}}$, which determine the power spectra of the spectral index fields. For these we impose narrower priors to exclude unphysically large power. Note that the data are still informative on the power spectra of the spectral index fields: the prior is wider than the typical posterior width, especially for the $B_{\mathrm{d}}$ and $B_{\mathrm{s}}$ parameters. 
When the model is applied to the PySM simulations in Sec.~\ref{sec:pysm_results}, we use a wider prior on the $A_{\mathrm{d}}$, $A_{\mathrm{s}}$, $\alpha_{\mathrm{d}}$, $\alpha_{\mathrm{s}}$, $\beta_{\mathrm{d}}$, and $\beta_{\mathrm{s}}$ parameters, see Table~\ref{table:prior}, to make sure the inferred posteriors lie within the supported parameter range.  

The CMB, dust, and synchrotron sky maps are generated from the prior draw using the forward model described in Sec.~\ref{sec:exp_setup}. 
For the pixelized maps, we use the HEALPix\footnote{http://healpix.sf.net} scheme with $N_{\mathrm{side}}=128$. 
For simplicity, we only simulate the $B$-mode component. This implies that even when a sky mask is applied to the resulting Stokes $Q$ and $U$ maps, there is no leakage from the $E$-mode signal to the $B$-mode estimate due the non-orthogonality of the $E$- and $B$-mode basis on the incomplete sky.\footnote{In reality, the additional variance in the $B$-mode estimate due to leaked $E$-mode signal is significant. However, there are several ways to solve this problem \citep{lewis_2003,smith_2006,bunn_2017}. For simplicity, we leave this detail to future work.}

Once the sky component maps are generated, the noisy and beam-convolved frequency maps are produced. We generate two sets, $\bm{d}^{(1)}$ and $\bm{d}^{(2)}$, of frequency maps that share signal but have independent noise realizations to mimic what would be done in a real analysis. See Appendix~\ref{app:instrument} for details on the noise and beams. We multiply the noise variance of each split by a factor two to incorporate the fact that each split is made from half of the observations. 
For the PySM tests in Sec.~\ref{sec:pysm_results}, we use a mask with an effective sky coverage of 21\%. This is roughly twice the sky fraction used for the SO forecasts \citep{so_forecast_2019, wolz_2024}. Our masked data will therefore include a larger variation of foreground properties and will provide a more challenging, and convincing, test case for our method. See Appendix~\ref{app:instrument} for details on the mask. For the other results we keep the data full sky. 

For the ILC-based compression step, we use the implementation of the constrained HILC and NILC algorithms from the \texttt{pyilc} python package \citep{mccarthy_2024}.\footnote{\url{https://ascl.net/2404.017}} 
For the NILC setup, we use four Gaussian needlets as the wavelet kernels in the $\mathbf{K}$ matrices in Eq.~\eqref{eq:nilc_cov}. The needlets are constructed following Eqs.~(38) and~(39) in \cite{mccarthy_2024} with FWHM parameters given by 300, 120, and $\SI{60}{\mathrm{arcmin}}$. The \texttt{ilc\_bias\_tol} parameter that determines the Gaussian smoothing of the covariance elements to suppress the ILC bias, see Eq.~(44) in \cite{mccarthy_2024}, is set to the default value of 0.01, yielding smoothing FWHM parameters of 122, 53, 28, and \SI{9}{\deg} for the four kernels. For the HILC setup, we use bands of $\Delta_{\ell} = 15$ to suppress the ILC bias. For the results in Sec.~\ref{sec:results_joint_nilc} we make use of alternative types of NILC compression; Table~\ref{table:nilc_types} describes the different setups. 

After the spectra are computed, see Eq.~\eqref{eq:cross_spectra}, they are binned by computing the mean of $\hat{C}_{\ell}$ in bins of width $\Delta \ell = 15$, between $30 \leq \ell \leq 200$. The length of the final data vector is 165. We find that this size is small enough for the current setup and postpone the exploration of further compression to future work.

The generation of each $\bm{\theta}$-$\bm{x}$ pair takes approximately ten CPU minutes. The vast majority of the time is spent in the NILC estimation. We expect  that with numerical optimization tuned to our setup the NILC evaluation costs can be reduced significantly, but we leave that to future work.

\subsection{Validation of the setup}\label{sec:validation}

\begin{figure}[t]
\includegraphics[width=\columnwidth]{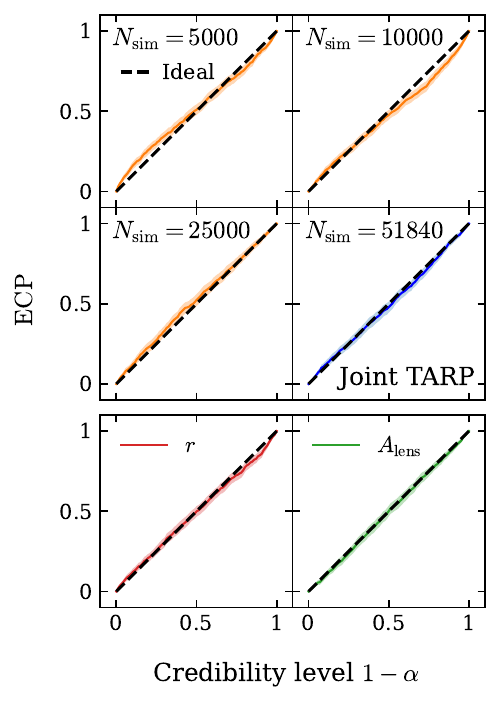}       
    \caption{TARP coverage tests. Top four panels: the joint coverage for different amounts of training simulations. Bottom two panels: the TARP test applied to the marginal posteriors of $r$ and $A_{\mathrm{lens}}$. The contours show $\pm2$ standard deviations of the test statistic estimated using bootstrap resampling.}\label{fig:tarp_comb}
\end{figure} 

We rely on coverage tests to assess the accuracy of the inference and the required number of simulations. These do not test the performance on out-of-distribution datasets, see Sec.~\ref{sec:pysm_results} for that, but verify whether the training set was sufficiently large and whether the normalizing flow architecture was sufficiently expressive.

Specifically, we use the TARP method by \cite{lemos_2023} to test the expected coverage probability ($\mathrm{ECP}$) of the posterior estimate. Expected coverage testing can be understood as checking whether the confidence contours, determined from the posterior estimator, on average contain the true parameters with the correct probability. More formally, given a set of $N$ parameter-data pairs $\{\bm{\theta}_i, \bm{x}_i \}_{i=1}^{N}$ drawn from the prior and a set of posterior samples generated from the posterior estimator $q$ for each prior sample: $\{ \hat{\bm{\theta}}_{ij} \}_{j=1}^{M}\ \forall \ i \in [1,N]$, the test amounts to determining the fraction $f_i$ of posterior samples that fall within a ball that is centered on $\bm{\theta}_{r,i}$ with radius extending to $\bm{\theta}_i$: 
\begin{align}
    f_i = \frac{1}{M}\sum_{j=1}^M \mathbbm{1} [d(\hat{\bm{\theta}}_{ij} , \bm{\theta}_{r,i} ) <  d(\bm{\theta}_{i} , \bm{\theta}_{r,i} )] \, .
\end{align}
Here, $d$ is a distance metric in parameter space, the Euclidean distance in our case, and $\bm{\theta}_{r,i}$ is a reference parameter vector that is drawn from an arbitrary reference sampling distribution $\tilde{P}(\bm{\theta}_{r,i} | \bm{x}_{i})$ that may depend on the data. \cite{lemos_2023} prove that the $\mathrm{ECP}$ associated with the spherical credible regions around the reference points $\bm{\theta}_{r,i}$ for a given credibility level $1-\alpha$ is given by the fraction of fractions $f_i$ that fall below $1 - \alpha$:
\begin{align}
    \mathrm{ECP}(\alpha) = \frac{1}{N}\sum_{i=1}^{N} \mathbbm{1}(f_i < 1-\alpha) \, .
\end{align}

In practice, this means that for an accurate posterior estimator, i.e.\ $q(\bm{\theta} | \bm{x}) = P(\bm{\theta} | \bm{x}) \ \forall \bm{\theta}, \bm{x} \sim P(\bm{\theta}, \bm{x})$, the TARP $\mathrm{ECP(\alpha)}$ estimate should match the credibility level  $1 - \alpha$. This should hold regardless of the reference distribution~$\tilde{P}$. By allowing $\tilde{P}$ to be conditioned on $\bm{x}_i$ the TARP test can avoid known failure modes of expected coverage tests based on the $\mathrm{ECP}$ associated with highest posterior density regions that only test whether an estimator is ``conservative''. Such tests, for example, do not detect the case where $q(\bm{\theta} | \bm{x}) = p(\bm{\theta})$ \citep{prangle_2013}.\footnote{\cite{lemos_2023} show how plots like Fig.~\ref{fig:tarp_comb} would look for overconfident, underconfident, and biased posterior estimators.}

In Fig.~\ref{fig:tarp_comb} we show the $\mathrm{ECP}$ determined from 864 posteriors computed from datasets created from parameters drawn from the prior. 
We apply the TARP test both to the joint posterior and to the individual marginal posteriors. Testing the joint distribution provides a global check that requires correct correlations between parameters in addition to coverage for the marginals. In contrast, testing the marginals ignores parameter correlations but can reveal cases where  different parameters offset each other in the joint coverage. 
Additionally, we show how the coverage depends on the amount of simulations used during training. The joint coverage appears to be effectively ideal, lying within the $2\sigma$ contours for all credibility levels.\footnote{A $1\sigma$ deviation in Fig.~\ref{fig:tarp_comb} does not imply that the posteriors are biased by one standard deviation. Instead, it conveys the confidence at which  incomplete coverage is detected with the 864 test trials.} More substantial deviations are seen when $\SI{5000}{}$ or $\SI{10000}{}$ simulations are used. For $\SI{25000}{}$ simulations, the coverage appears sufficiently complete. As a conservative choice, we approximately double the amount of simulations ($N_{\mathrm{sim}} = \SI{51840}{}$) for the results in this paper. 
The tests for the individual marginal posteriors, shown in the bottom panel of Fig.~\ref{fig:tarp_comb} for $r$ and $A_{\mathrm{lens}}$ and Fig.~\ref{fig:tarp_marg_fg} for the foreground parameters, reveal slightly larger deviations, including a hint of overconfidence for~$r$. While these tests suggest a small degree of incomplete coverage for certain parameters, we judge the deviations too small to impact the results of the paper meaningfully. 
In Fig.~\ref{fig:tarp_emoped} we demonstrate that the joint TARP results are insensitive to the reference distribution $\tilde{P}$: we compare a uniform distribution over the (hyperrectangular) support of the prior draws, which is the default used for the other tests, to $\tilde{P}(\bm{\theta}_{r,i} | \bm{x}_{i}) = g(\bm{x}_i) + \epsilon_i$, where $g : \mathbb{R}^{N_{\bm{x}}} \rightarrow \mathbb{R}^{N_{\bm{\theta}}}$ is the e-MOPED compression\footnote{Estimated with the empirical covariance matrix of 864 additional data vectors created from parameters drawn from the prior.} from \cite{park_2025} and $\epsilon_i$ a Gaussian variate with standard deviation given by 1\% of the support of the prior draws.   
Overall, the coverage tests indicate that our training setup and normalizing flow architecture provide accurate inference. We make suggestions for further improvements in Sec.~\ref{sec:discussion}.

\section{Results} 

\subsection{SBI compared to the multi-frequency likelihood}

\begin{figure}[t]
    \centering \includegraphics[width=\columnwidth]{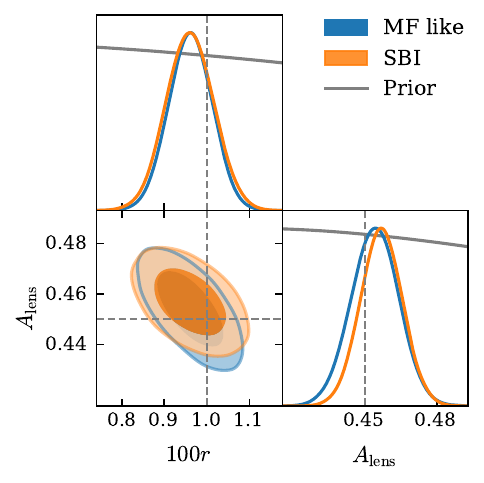}
    \caption{Comparison of posteriors from the standard multi-frequency power spectrum likelihood (blue) and our new joint-NILC SBI technique (orange). Both use the same data, generated without spatial variation in the spectral index fields or other sources of statistical anisotropy, making the likelihood approach statistically optimal; this comparison thus validates that our method reproduces the correct results in this simple setting.}\label{fig:mcmc_cosmo_only}
\end{figure} 

\begin{figure*}[t]
    \centering \includegraphics[width=\textwidth]{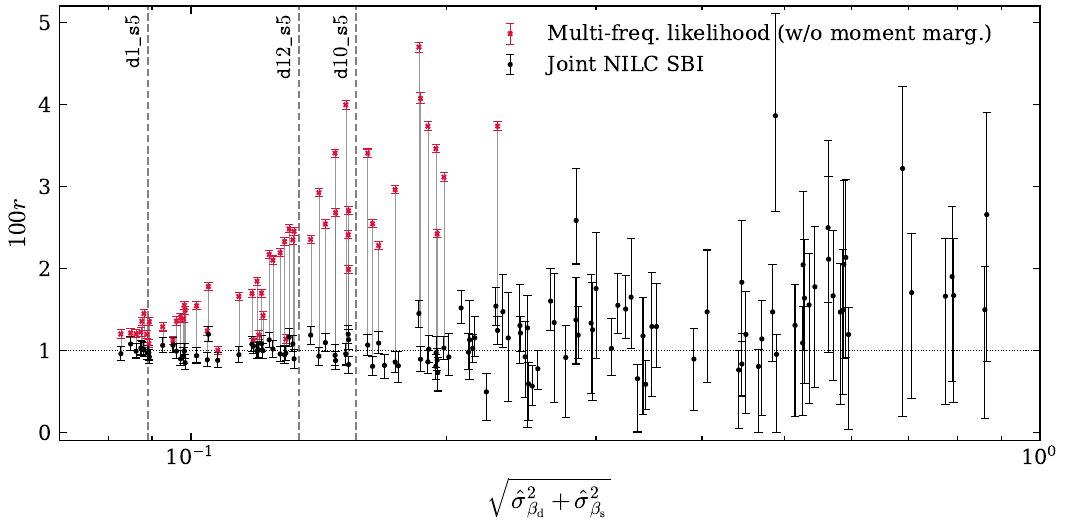}       
    \caption{Comparison between inferred $r$ values for our new joint NILC SBI method and the standard multi-frequency power-spectrum-likelihood-based method (i.e.\ without the moment marginalization of \cite{azzoni_2021}). The error bars show the highest density interval that contains 68\% of the probability. The points are ordered based on the mean standard deviation of the $\beta_{\mathrm{d}}$ and $\beta_{\mathrm{s}}$ fields that is inferred by the SBI method. We find this to be a decent indicator of both the bias of the likelihood method and the posterior width of the SBI method. The test data for this plot are generated for fixed $r=0.01$. The SBI results produce unbiased results regardless of spectral index power, while the likelihood-based method is biased over the entire range and becomes particularly biased as the inferred spectral index power increases. Solid vertical gray lines connect SBI and likelihood-based points that are generated from the same input data. Dashed solid lines show the standard deviation of $\beta_{\mathrm{d}}$ and $\beta_{\mathrm{s}}$ for three different combinations of PySM models evaluated over a 21\% sky mask, see Sec.~\ref{sec:pysm_results}. For the $\texttt{d12}$ model, $\sigma_{\beta_\mathrm{d}}$ has been determined from a per-pixel fit of Eq.~\eqref{eq:dust_sed} and should be considered a conservatively low estimate. To aid the visualization, we have thinned the distribution of points for this plot by randomly selecting ten points in each of 15 logarithmic bins. In addition, likelihood points at high values of $(\hat{\sigma}_{\beta_\mathrm{d}}^2 + \hat{\sigma}_{\beta_\mathrm{s}}^2)^{1/2}$ for which the power-spectrum-based MCMC sampling fails due to multimodality are not displayed.}\label{fig:mean_r_mcmc}
\end{figure*} 

We first compare the results from our SBI setup to the standard multi-frequency power-spectrum-based likelihood. The test data for this comparison follow the data model described in Sec.~\ref{sec:sim_setup}, which was also used to train the NPE. As shown in Sec.~\ref{sec:validation}, our setup produces accurate posteriors for data vectors that are consistent with the assumed data model. Still, the comparison to the likelihood-based method serves two purposes: (i)~as an additional sanity check: since the likelihood-based method is statistically optimal when the spectral indices are constant, i.e.\ $B_{\mathrm{d}} = B_{\mathrm{s}} = \gamma_{\mathrm{d}} = \gamma_{\mathrm{s}} =0$, it will reveal whether our method is similarly optimal in this limit; and (ii),~to reveal how the likelihood-based method becomes biased once the spectral indices vary spatially, while the NPE posteriors remain unbiased.

We implement the multi-frequency power spectrum likelihood following the description in \cite{azzoni_2021}. The likelihood includes the $r$, $A_{\mathrm{lens}}$, $A_{\mathrm{d}}$, $\alpha_{\mathrm{d}}$, $\beta_{\mathrm{d}}$, $A_{\mathrm{s}}$, $\alpha_{\mathrm{s}}$, $\beta_{\mathrm{s}}$, and $\rho_{\mathrm{ds}}$ parameters from the forward model described in Sec.~\ref{sec:exp_setup}. Our implementation does not include the moment method from \cite{azzoni_2021}, which would be needed to sample $B_{\mathrm{d}}$, $\gamma_{\mathrm{d}}$, $B_{\mathrm{s}}$, and $\gamma_{\mathrm{s}}$. Broadly speaking, the implemented method is the same as used by \cite{bicepkeck_2021}. Details on the implementation and MCMC sampling are given in Appendix~\ref{app:mflike}.

Fig.~\ref{fig:mcmc_cosmo_only} compares estimates for the marginal posterior on $r$ and $A_{\mathrm{lens}}$ from both methods. The true values for this test are given by: $r=0.01$, $A_{\mathrm{lens}} = 0.45$, $A_{\mathrm{d}} = \SI{29}{\micro \kelvin \squared}$, $\alpha_{\mathrm{d}} = -0.3$, $\beta_{\mathrm{d}} = 1.55$, $A_{\mathrm{s}} =  \SI{1.5}{\micro \kelvin \squared}$, $\alpha_{\mathrm{s}} = -0.6$, $\beta_{\mathrm{s}} = -2.8$, and $\rho_{\mathrm{ds}} = 0.1$. We evaluate the analytic covariance matrix used in the likelihood method at these true parameters. For this test, the NPE has been trained on simulations with $B_{\mathrm{d}} = B_{\mathrm{s}} = \gamma_{\mathrm{d}} = \gamma_{\mathrm{s}} =0$ to provide a fair comparison to the likelihood-based method, which assumes the data model without spatially varying spectral indices. 
The estimates are in excellent agreement, which implies that the SBI setup is effectively statistically optimal in the limit of isotropic $\beta_{\mathrm{d}}$ and $\beta_{\mathrm{s}}$. 
Small differences, such as seen in the $A_{\mathrm{lens}}$ marginal for this realization, are not unexpected given the different handling of the foregrounds, see Appendix~\ref{app:full_mf_comparison} for more details. 
We emphasize that although the training and test data are generated without spatially varying $\beta_{\mathrm{d}}$ and $\beta_{\mathrm{s}}$, the compression used for this test still uses the five-component joint NILC ($\texttt{c+d+s+dbd+dbs}$) compression, see Table~\ref{table:nilc_types}. This confirms that our data vector size (165) is small enough for the normalizing flow to produce optimal estimates of the cosmological parameters.\footnote{Fig.~\ref{fig:fig:mcmc_full} shows the posterior of all nine parameters of the forward model used for this SBI setup. The NPE marginal posteriors of the synchrotron parameters are slightly broader. We attribute this to two factors: (i) NILC being slightly less optimal due to the limited angular resolution of the needlets, slightly suboptimal fiducial SED parameters, and the sample variance in its empirical covariance matrix (whereas the likelihood is using the exact analytic covariance matrix evaluated at the true parameters), and (ii) incomplete convergence of the NPE, which is likely the dominant effect. We return to this point in the discussion in Sec.~\ref{sec:discussion} and Appendix~\ref{app:full_mf_comparison}.}

In Fig.~\ref{fig:mean_r_mcmc}, we show how the inferred values of $r$ change with increased inferred spatial variation in the spectral index fields. 
To isolate the effect of SED anisotropies, the test data for this plot are generated from 864 sets of input parameters drawn from a modified prior that keeps the prior distribution from Table~\ref{table:prior} for the parameters that govern the anisotropy of the foreground SED ($B_{\mathrm{d}}$, $B_{\mathrm{s}}$, $\gamma_{\mathrm{d}}$ and $\gamma_{\mathrm{s}}$) and fixes the other parameters to the following values: $r=0.01$, $A_{\mathrm{lens}} = 0.5$, $A_{\mathrm{d}} = \SI{28}{\micro\kelvin\squared}$, $\alpha_{\mathrm{d}} = -0.3$,  $\beta_{\mathrm{d}} = 1.55$, $A_{\mathrm{s}} = \SI{1.5}{\micro\kelvin\squared}$, $\alpha_{\mathrm{s}} = -0.7$,  $\beta_{\mathrm{s}} = -2.9$, and $\rho_{\mathrm{ds}}~=~0.1$.
For convenience, we summarize the spatial variation, which depends on both $B$ and $\gamma$ parameters, as a single variance number:
\begin{align}\label{eq:beta_var}
    \sigma^2_{\beta_{\mathrm{d}}} = \frac{1}{4 \pi}\sum_{\ell=2}^{200} (2 \ell + 1) \, C^{\beta_{\mathrm{d}}}_{\ell}(B_{\mathrm{d}}, \gamma_{\mathrm{d}}) \, ,
\end{align}
where $C^{\beta_{\mathrm{d}}}_{\ell}(B_{\mathrm{d}}, \gamma_{\mathrm{d}})$ is given by Eq.~\eqref{eq:gamma_dust}. For the inferred variance, $\hat{\sigma}^2_{\beta_{\mathrm{d}}}$, in Fig.~\ref{fig:mean_r_mcmc} the posterior samples of $B_{\mathrm{d}}$ and $\gamma_{\mathrm{d}}$ are used. The analogous quantity is computed for the synchrotron case. 
We compare results from our new SBI method with those from the standard multi-frequency power-spectrum-based likelihood method (i.e.\ without the moment marginalization of \cite{azzoni_2021}). As expected \citep{azzoni_2021,liu_2025}, the likelihood-based posteriors are biased when such a spatial dependence is introduced. In our setup, the bias is so severe that the standard likelihood-based method appears to be effectively unusable,\footnote{In fact, for large enough spatial variation, the mismatch between the data model assumed by the likelihood-based approach and the data causes multiple modes to appear in the posterior. As our MCMC sampler cannot reliably sample such distributions, we do not display points in Fig.~\ref{fig:mean_r_mcmc} for which we detect multimodality.} while the SBI results remain unbiased. 
As the SBI method infers a larger variation in the spectral indices, it adjusts the uncertainty on $r$ accordingly. We explore this behavior further in the next section.

\subsection{Information content of the joint NILC compression}\label{sec:results_joint_nilc}

\begin{figure*}[t]
    \centering \includegraphics[width=\textwidth]{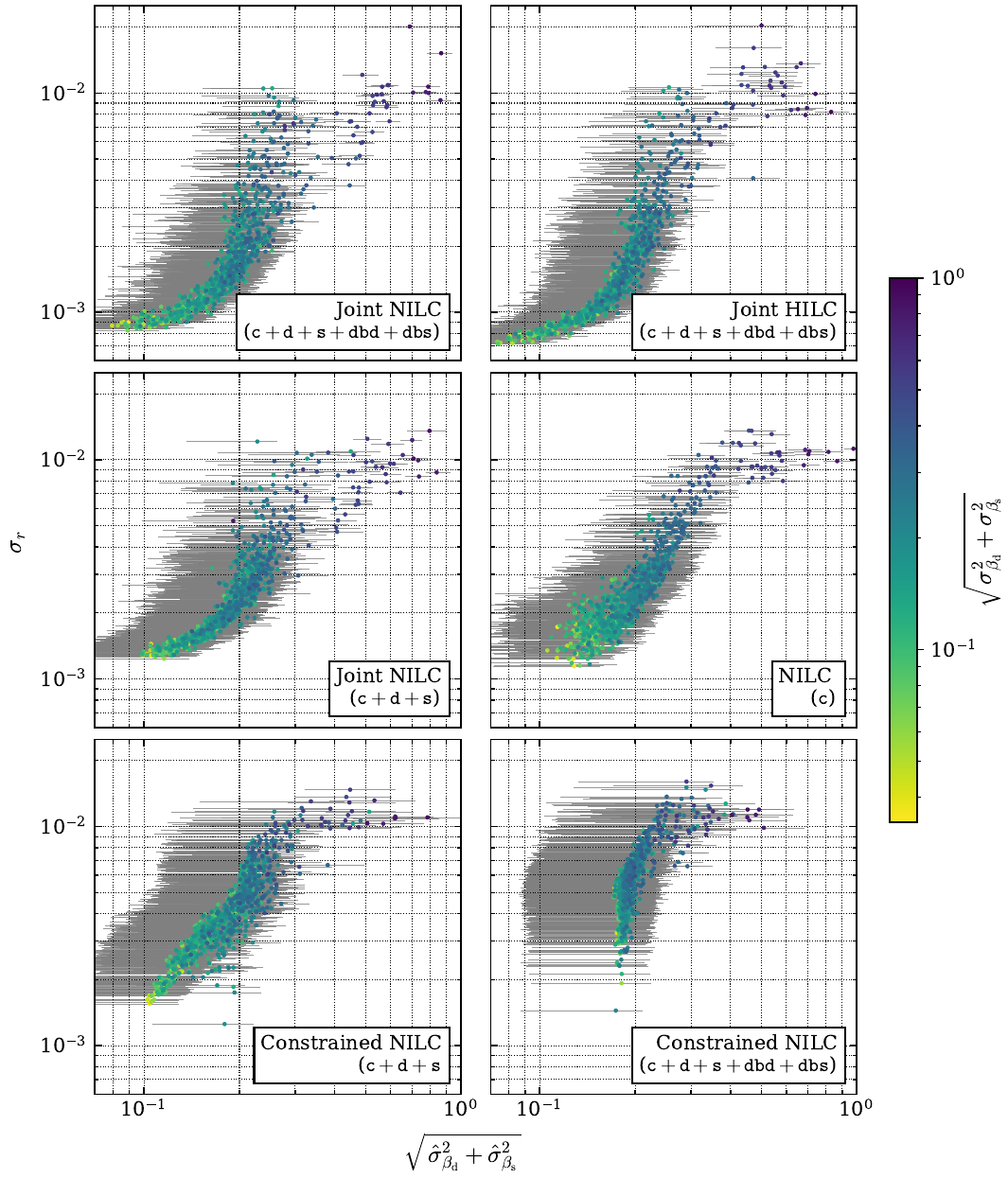}       
    \caption{Standard deviation (half of width of 68\% highest density interval) of $r$ as function of the mean inferred standard deviation of the $\beta_{\mathrm{d}}$ and $\beta_{\mathrm{s}}$ fields. Colors indicate the true value of the standard deviation. The horizontal gray error bars denote the 68\% highest density interval of the posterior on $(\sigma_{\beta_{\mathrm{d}}}^2 + \sigma_{\beta_{\mathrm{s}}}^2)^{1/2}$. Each dot summarizes the posterior for one of the simulated datasets described in Sec.~\ref{sec:results_joint_nilc} and the panels show the different SBI setups described in Table~\ref{table:nilc_types}. It is clear how the inclusion of foreground maps improves the constraining power of the joint NILC method compared to the standard NILC and constrained NILC setups. The inferred power of the spectral index fields appears to be a good indicator for the uncertainty on $r$.}\label{fig:sigma_r}
\end{figure*} 

In this section we take a closer look at the information content of the compressed data vectors. We compare our proposed joint NILC (\texttt{CMB+d+s+dbd+dbs}) setup with four alternative compression schemes, see Table~\ref{table:nilc_types}, which use fewer or no foreground maps. 

Fig.~\ref{fig:sigma_r} shows how the posterior width on $r$ varies with the inferred standard deviation of the spectral index fields (see Eq.~\eqref{eq:beta_var}) for each of the five different compression schemes. Our new joint-NILC SBI method produces the lowest-variance estimate on $r$. The clear relationship between $\sigma_r$ and the standard deviation of the spectral index fields is notable. The fidelity of the $r$ estimate appears closely linked to the ability to measure the spatial variations in the foreground SEDs. The importance of including the spectra involving the $\delta \beta_{\mathrm{d}}$ and $\delta \beta_{\mathrm{s}}$ estimates in the compression is clear when comparing to the joint NILC ($\texttt{c+d+s}$) and ($\texttt{c}$) schemes, which compress the frequency maps into only three or one sky component map. As a consequence, these compressed data vectors contain little information on the SED anisotropy of a particular foreground realization. The inference thus has to fall back on the average assumed anisotropy suggested by the prior and cannot achieve the sensitivity on $r$ that our fiducial joint-NILC method shows. The two constrained NILC cases, which similar to the NILC ($\texttt{c}$) scheme compress the data into a single CMB estimate, perform worst. Compared to the NILC ($\texttt{c}$) scheme, the deprojection of the sky components increases the variance of the CMB estimate, which makes it even more difficult to extract remaining information about the contamination due to SED anisotropy compared to the NILC ($\texttt{c}$) case.
For context, evaluated at the true $(\sigma^2_{\beta_{\mathrm{d}}} + \sigma^2_{\beta_{\mathrm{s}}})^{1/2}$ of the medium-complexity $\texttt{d10\_s5}$ PySM foreground model (see Fig.~\ref{fig:mean_r_mcmc} and Section~\ref{sec:pysm_results}), the increase in $\sigma_r$ compared to our proposed method is $1.2$, $2.3$, and $3.1$ for NILC~($\texttt{c}$), constrained NILC ($\texttt{c+d+s}$), and constrained NILC ($\texttt{c+d+s+dbd+dbs}$), respectively. For the low-complexity $\texttt{d1\_s5}$ foreground model the differences are even larger: $1.5$, $2.6$, and $4.1$.

In Fig.~\ref{fig:sigma_r} we also show results for the joint HILC-based compression. This approach produces narrower $r$ posteriors for cases with low spatial power in the spectral index fields. The NILC-based compression performs better for the most extreme values of $(\sigma^2_{\beta_{\mathrm{d}}} + \sigma^2_{\beta_{\mathrm{s}}})^{1/2}$. 
Comparing $\sigma_{r}$ on a realization-by-realization basis, we find that the HILC-based compression outperforms NILC by roughly 5-10\% for $(\sigma^2_{\beta_{\mathrm{d}}} + \sigma^2_{\beta_{\mathrm{s}}})^{1/2} < 0.1$, while for $(\sigma^2_{\beta_{\mathrm{d}}} + \sigma^2_{\beta_{\mathrm{s}}})^{1/2} > 0.4$, the NILC-based compression performs better by roughly 20\%. For intermediate values, the two methods perform roughly equally in terms of $\sigma_r$. 
The decreased variance for nearly isotropic spectral index maps is a consequence of the lower sample variance of the HILC method, while in the regime of high SED anisotropy, the spatial localization of the NILC covariance matrix is better suited to the non-Gaussian structure in the maps. In the next section, we show how this localization yields a more robust $r$ estimate for realistic foreground maps compared to HILC.

\subsection{Applied to PySM foreground simulations}\label{sec:pysm_results}

\begin{figure*}[htbp]
    \centering \includegraphics[width=\textwidth]{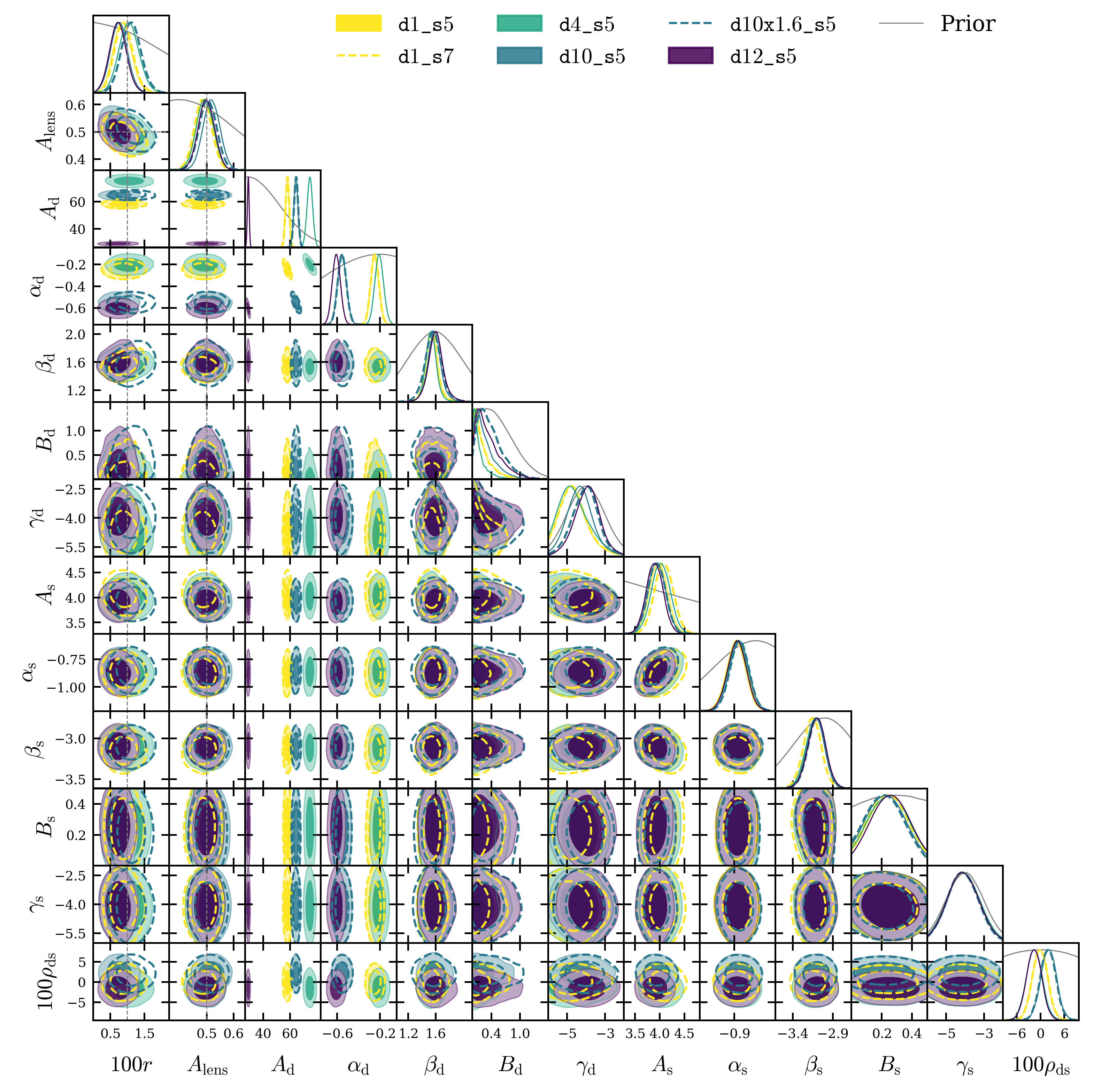}       
    \caption{Posterior distributions for different PySM foreground simulations. The CMB and noise realizations are shared between all cases. The true values of the $r$ and $A_{\mathrm{lens}}$ parameters are denoted with dashed gray lines. It can be seen that the same conditional normalizing flow, trained on the relatively simple simulations used in the work, produces an unbiased posterior for the $r$ and $A_{\mathrm{lens}}$ parameters for all types of foreground models, demonstrating the robustness provided by the NILC-based compression strategy. The marginal posteriors of $B_{\mathrm{d}}$ and $\gamma_{\mathrm{d}}$ indicate that the method is able to infer that the more complex models have increased power in the dust spectral index field.}\label{fig:posteriors_pysm}
\end{figure*}

\begin{figure}[h]
\includegraphics[width=\columnwidth]{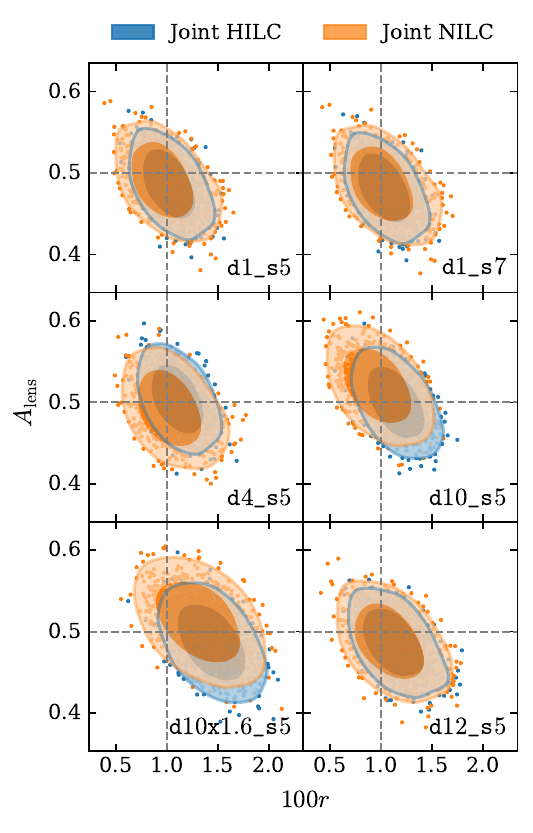}       
    \caption{The distribution of posterior means on $r$ and $A_{\mathrm{lens}}$ for 864 simulated datasets that share the same PySM foreground realization but vary the noise and CMB realizations. The panels show the different combinations of PySM models and colors denote whether the compression was done using the joint HILC ($\texttt{c+d+s+dbd+dbs}$) method (blue) or the joint NILC ($\texttt{c+d+s+dbd+dbs}$) method (orange). Contours enclose 68\% and 95\% of the distribution of mean values; the contours should not be interpreted as posteriors on $r$ and $A_{\mathrm{lens}}$. It can be seen that the NILC method remains unbiased (at the 68\% level or better) for all foreground models while the distributions for the HILC method are less well centered on the true value and show a bias for the most complex foreground model ($\texttt{d10x1\!.\!6\_s5}$). This demonstrates that the anisotropic nature of the NILC covariance matrix improves the robustness to model misspecification.}\label{fig:pysm_bias}
\end{figure} 

\begin{figure}[h]
\includegraphics[width=\columnwidth]{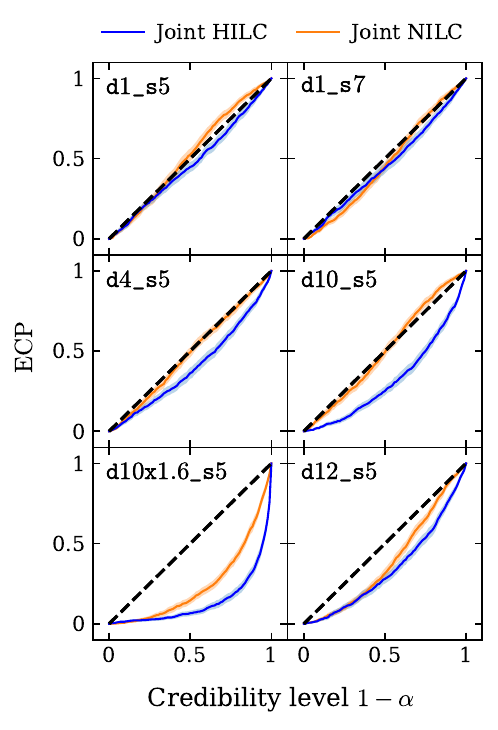}       
    \caption{TARP coverage tests of the $r$ marginal posterior for the joint HILC ($\texttt{c+d+s+dbd+dbs}$) method (blue) and the joint NILC ($\texttt{c+d+s+dbd+dbs}$) method (orange). The panels show the different combinations of PySM models. The contours indicate $\pm2$ standard deviations of the test statistic estimated using bootstrap resampling. The dashed gray line shows the coverage for a perfectly accurate posterior estimator. An $\mathrm{ECP}$ value consistently below the dashed line is indicative of a biased posterior \citep{lemos_2023}. While the test clearly detects a bias for the $\texttt{d10x1.6\_s5}$ and $\texttt{d12\_s5}$ models, even for the joint NILC compression, the biases, for input $r=0.01$, are quite modest, $1.1\sigma$ and $0.3\sigma$, respectively.}\label{fig:pysm_tarp}
\end{figure}

\begin{figure*}[htbp]
    \centering \includegraphics[width=\textwidth]{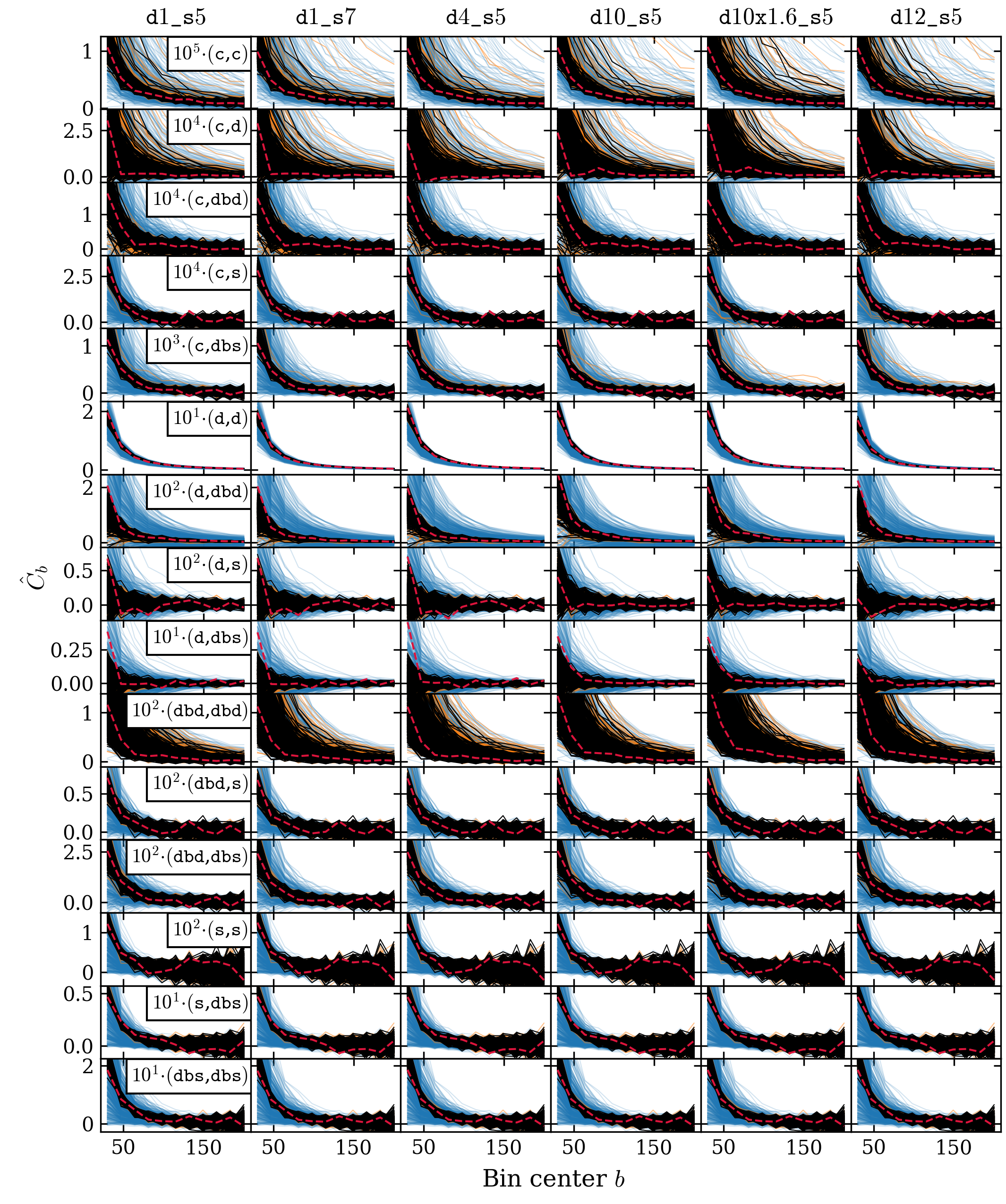}       
    \caption{Comparison of the compressed data vectors of the PySM test datasets (dashed red) to data vectors corresponding to the prior (blue) and 864 samples from the posterior predictive distribution (68\% highest posterior density regions in black, remainder in orange). The test sets and posteriors are the same as shown in Fig.~\ref{fig:posteriors_pysm}. Each column represents a different combination of PySM models, while each row shows one of the spectra that make up the compressed data vector. The spectra are labeled by the pair of estimated sky components that are correlated: $\mathtt{c}$: CMB, $\mathtt{d}$: dust, $\mathtt{s}$: synchrotron, $\mathtt{dbd}$: $\delta \beta_{\mathrm{d}}$, and $\mathtt{dbs}$:~$\delta \beta_{\mathrm{s}}$.}\label{fig:data_pysm}
\end{figure*} 

\begin{figure}[t]
\includegraphics[width=\columnwidth]{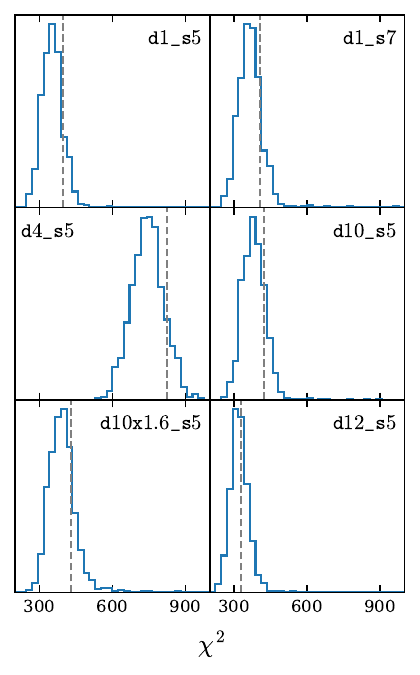}       
    \caption{Posterior predictive distribution of $\chi^2$ values compared to the $\chi^2$  value of the observed data (gray dashed line) for each PySM combination.}\label{fig:chi2_test}
\end{figure} 

Finally, we turn to the most important question: does the inference still work when applied to realistic foreground simulations that do not follow the simple data model assumed for the training set? To test this we apply our proposed NILC SBI setup to the PySM simulations \citep{thorne_2017,zonca_2021,pan_exp_2025}.

We use the following PySM models: \texttt{d1}, \texttt{d4}, \texttt{d10}, \texttt{d12} for dust and \texttt{s5} and \texttt{s7} for synchrotron. These simulations are more complex  than the simulations used to train the NPE in several ways: (i) statistical anisotropy of the dust and synchrotron amplitudes, the spectral indices, and the dust temperature; (ii) non-Gaussianity in all previously mentioned fields; (iii) non-Gaussian and anisotropic correlations between all previously mentioned fields \citep{liu_2025}; and (iv) deviations from the assumed functional forms of the foreground SEDs. Specifically, the $\texttt{d12}$ model contains line-of-sight variations that distort the SED slightly away from the modified blackbody functional form, the $\texttt{d4}$ model contains a spatially varying mixture of two modified blackbodies, and the $\texttt{s7}$ model has a non-power-law synchrotron SED. We additionally construct a more complex version of the $\texttt{d10}$ dust model, the $\texttt{d10x1.6}$ model, for which the anisotropic part of the spectral index field has been multiplied by a factor 1.6. The isotropic part, defined as the average within the sky mask, is kept unchanged. For this rescaled model, the dust temperature is set to its average within the mask to avoid lowering the amount of frequency decorrelation due to the spatial anti-correlation between the spectral index and temperature fields, as explained in \cite{liu_2025}.

We retrain the model using the widened prior listed in Table~\ref{table:prior} and apply the sky mask as part of the forward model. See Sec.~\ref{sec:sim_setup} for details. The test data for this section are produced by replacing the foreground realization with the PySM models evaluated at our frequency bands. We consider six combinations of dust and synchrotron models: $\texttt{d1\_s5}$, $\texttt{d1\_s7}$, $\texttt{d4\_s5}$, $\texttt{d10\_s5}$, $\texttt{d12\_s5}$, and $\texttt{d10x1.6\_s5}$.\footnote{We produce the PySM maps at their native resolution ($N_{\mathrm{side}}=512$ or $2048$ depending on the model), apply an apodized mask to the brightest parts of the Galactic plane to avoid ringing, transform to harmonic space, band-limit to $\ell_{\mathrm{max}} = 200$, and  convert the $B$-mode coefficients to $N_{\mathrm{side}} = 128$ $Q$ and $U$ maps. This downgrading introduces a type of SED distortion from averaging pixels with different spectra \citep{chluba_2017}, which is absent in the simulations used to train the NPE. We retain this mismatch, as any viable inference pipeline should be able to handle such effects.} For each combination, we produce 864 datasets with different CMB and noise realizations. All other aspects of the forward model, such as the beams and noise, are unchanged. We emphasize that a single NPE model is applied to all different test datasets; there has been no tuning to fit individual PySM models nor has the sky mask been tuned. In this sense, the results presented here are from a blind analysis.

In Fig.~\ref{fig:posteriors_pysm}, we compare the posteriors obtained for each combination of dust and synchrotron PySM models. The CMB and noise realizations are identical across all cases. For each model, the $r$ and $A_{\mathrm{lens}}$ posteriors convincingly overlap with the input values. The mean inferred $r$ value varies slightly between models, but it is difficult to assess the significance from a single posterior; this will be explored in the next test. The marginal posteriors on $B_{\mathrm{d}}$ and $\gamma_{\mathrm{d}}$ reveal a hierarchy in the complexity of the dust models. Ordered by the inferred anisotropy in the spectral index field, we have: $\texttt{d4}$, $\texttt{d1}$, $\texttt{d10}$, $\texttt{d12}$, and $\texttt{d10x1.6}$. The $\texttt{d10x1.6}$ and $\texttt{d12}$ values are comparable. This ordering matches our expectation: while $\texttt{d4}$ uses anisotropic templates for its two dust temperature fields, it uses two isotropic  $\beta_{\mathrm{d}}$ templates and thus should show little spatial SED variations in our frequency bands \citep{thorne_2017}. The GNILC-based $\beta_{\mathrm{d}}$ template used by the $\texttt{d10}$ model is more spatially varying than the Commander-based template used for the $\texttt{d1}$ model \citep{pan_exp_2025}. Finally, 
while a per-pixel fit  of Eq.~\eqref{eq:dust_sed} ($N_{\mathrm{side}}=128$) reveals that the $\texttt{d12}$ model has a $\sigma_{\beta_{\mathrm{d}}}$ that is comparable to the $\texttt{d10}$ model in our patch, the SED distortions due to anisotropic line-of-sight variations included in $\texttt{d12}$ should increase the power of the SED anisotropies compared to $\texttt{d10}$. Regarding synchrotron SED variations, the posteriors show that the compressed data contain information on the $B_{\mathrm{s}}$ parameter, whereas the $\gamma_{\mathrm{s}}$ parameter remains prior-dominated.

While the results for the PySM models follow the expected hierarchy in terms of the complexity of the $\beta_{\mathrm{d}}$ field, the $B_{\mathrm{d}}$ marginal posteriors in Fig.~\ref{fig:posteriors_pysm} might seem concerning, as they do not significantly depart from zero. In Appendix~\ref{app:pysm_sbd}, we show that nonzero power in the $\beta_{\mathrm{d}}$ field is clearly detected when the $B_{\mathrm{d}}$ and $\gamma_{\mathrm{d}}$ parameters are transformed to the standard deviation of the field using Eq.~\eqref{eq:beta_var}. The appearance of the $B_{\mathrm{d}}$ marginal is due to a mild projection, or ``prior volume'', effect: 
as $B_{\mathrm{d}}$ approaches zero, the data become compatible with all $\gamma_{\mathrm{d}}$ values supported by the prior, causing the contours in the $B_{\mathrm{d}}$-$\gamma_{\mathrm{d}}$ plane to expand at low $B_{\mathrm{d}}$. This increase in parameter-space area leads to the excess of probability at low values in the marginal of $B_{\mathrm{d}}$, giving the impression of a non-detection despite the detection of significant power. This effect is a well-known consequence of power-law parameterizations. In future work it would be interesting to explore a different parameterization of the spectral index power or more sophisticated priors on $B_{\mathrm{d}}$ and $\gamma_{\mathrm{d}}$. 

\begin{deluxetable}{c|ll}
\tablecaption{Constraints on $100r$ for the PySM-based test sets\label{table:pysm_results}}
\tablehead{PySM combination & \colhead{Joint HILC} & \colhead{Joint NILC} 
}
\startdata
    \texttt{d1\_s5} &$1.02 \pm 0.19$&$0.97 \pm 0.26$\\
    \texttt{d1\_s7}&$1.03 \pm 0.19$&$1.02 \pm 0.26$ \\
    \texttt{d4\_s5} &$1.11 \pm 0.21$&$1.03 \pm 0.28$\\
    \texttt{d10\_s5}&$1.15 \pm 0.21$&$0.97 \pm 0.26$\\
    \texttt{d10x1.6\_s5}&$1.44 \pm 0.25$&$1.32 \pm 0.30$\\
    \texttt{d12\_s5}&$1.13 \pm 0.21$&$1.09 \pm 0.27$
\enddata
\tablecomments{Posterior means and widths for the 864 test datasets with input $r$ set to $0.01$, as used for Fig.~\ref{fig:pysm_bias}. The reported errors are given by half of the 68\% highest posterior density interval.  Means and errors are averaged over the test datasets.}
\end{deluxetable}

Fig.~\ref{fig:pysm_bias} shows the distribution of posterior means for the $r$ and $A_{\mathrm{lens}}$ parameters for the 864 different CMB and noise realizations. Additionally, Table~\ref{table:pysm_results} reports the $r$ posterior means and widths averaged over these test sets. The test data are all drawn from a modified prior with $r=0.01$ and $A_{\mathrm{lens}} = 0.5$. Our proposed joint NILC method produces unbiased results for each of the six PySM combinations. The true parameter values lie at the border of the 68\% interval of the distribution for the $\texttt{d10x1.6\_s5}$ combination. All other combinations lie comfortably within the 68\% interval. The figure additionally shows results for the joint HILC compression. Although the contours are slightly narrower, the results are biased for the $\texttt{d10x1.6\_s5}$ combination and generally are less centered on the true value, revealing that the HILC method, with its non-spatially-localized covariance matrix, is less robust to the mismodeling between our forward model and the PySM simulations. 

In Fig.~\ref{fig:pysm_tarp} we show a more stringent bias test using the TARP method from Sec.~\ref{sec:validation}, applied to 864 test datasets based on the PySM simulations but with varying CMB and noise. Unlike the test set used for Fig.~\ref{fig:pysm_bias}, the $r$ and $A_{\mathrm{lens}}$ parameters are not fixed, but drawn from the prior. The results reveal unbiased posteriors for the $\texttt{d1\_s5}$, $\texttt{d1\_s7}$, $\texttt{d4\_s5}$, and $\texttt{d10\_s5}$ combinations, while the $\texttt{d12\_s5}$ and, especially, the $\texttt{d10x1.6\_s5}$ models show biases. 
Although the TARP test detects biases for the joint NILC method applied to the two most complex models, the results from Fig.~\ref{fig:pysm_bias} and inspection of the individual posteriors
suggest that the biases themselves are small: substantially below one standard deviation of the posterior width for the $\texttt{d12\_s5}$ combination and slightly below one for $\texttt{d10x1.6\_s5}$. The result highlights the sensitivity of the TARP test to small biases and the practical benefit of using the NPE method to rapidly generate large numbers of posterior distributions.
Similar to Fig.~\ref{fig:pysm_bias}, the joint HILC compression performs consistently worse in Fig.~\ref{fig:pysm_tarp}, showing biases already for the $\texttt{d4\_s5}$ and $\texttt{d10\_s5}$ models. 
This implies that any purely harmonic method will suffer from similar (or worse) bias. In particular, given that the joint HILC compression is effectively equivalent to a multi-frequency power-spectrum-based likelihood method (see \cite{surrao_2024} for a detailed discussion), our joint HILC results should apply to the moment expansion method from \cite{azzoni_2021}.

We illustrate how for a real analysis, with a single observed data vector, posterior predictive checks \citep{gelman_1996} can be used to look for signs of mismodeling. The posterior predictive distribution is the distribution of replicated data $\bm{x}_{\mathrm{r}}$ conditioned on the observed data $\bm{x}_{\mathrm{obs}}$:
\begin{align}
    P(\bm{x}_{\mathrm{r}} | \bm{x}_{\mathrm{obs}}) = \int P(\bm{x}_{\mathrm{r}} | \bm{\phi})P(\bm{\phi} | \bm{\theta})P(\bm{\theta} | \bm{x}_{\mathrm{obs}}) \mathrm{d}\bm{\phi} \mathrm{d} \bm{\theta} \, .
\end{align}
We have re-introduced the vector of latent parameters $\bm{\phi}$ from Eq.~\eqref{eq:forward_model_gen}. In Fig.~\ref{fig:data_pysm}, we take the same datasets that were used for Fig.~\ref{fig:posteriors_pysm} and compare them to samples from the posterior predictive distribution and the prior predictive distribution:
\begin{align}
    P(\bm{x}_{\mathrm{r}}) = \int P(\bm{x}_{\mathrm{r}} | \bm{\phi})P(\bm{\phi} | \bm{\theta})P(\bm{\theta}) \mathrm{d}\bm{\phi} \mathrm{d} \bm{\theta} \, .
\end{align}
The overall consistency between the data vectors and the samples from the posterior predictive distribution further indicates that our forward model describes the PySM-based data after compression with sufficient accuracy. Any remaining discrepancy suggests either a limitation of the model or, from the compression perspective, a failure to suppress the discrepancy between simulations and data. Curiously, we do find small deviations: the first multipole bin of the dust-$\delta \beta_{\mathrm{s}}$ ($\mathtt{d}, \mathtt{dbs}$) power spectrum lies clearly outside the distribution, especially for the $\texttt{d1\_s5}$, $\texttt{d1\_s7}$, and $\texttt{d4\_s5}$ combinations. Since our model does not include any intrinsic correlation between dust or synchrotron amplitudes and their spectral indices, it is unable to capture such correlation if they truly exist in some of the PySM combinations. For now, the deviations appear small enough to not impact the inference of $r$, but this would be interesting to explore in future work.

Fig.~\ref{fig:chi2_test} presents a posterior predictive check that uses the  $\chi^2$ statistic. The statistic is defined as $\chi^2 : \mathbb{R}^{N_{\bm{x}}} \rightarrow \mathbb{R}_{>0}$ with $\chi^2 = \bm{x}^{\top} \mathbf{C}^{-1}\bm{x}$, with an $N_{\bm{x}} \times N_{\bm{x}}$ covariance matrix $\mathbf{C}$ that is estimated from a separate set of 864 simulations.\footnote{The modified prior used to generate these simulations uses fixed parameters: $r=0.01$, $A_{\mathrm{lens}} = 0.5$, $A_{\mathrm{d}} = \SI{50}{\micro\kelvin\squared}$, $\alpha_{\mathrm{d}} = -0.3$,  $\beta_{\mathrm{d}} = 1.55$, $B_{\mathrm{d}} = 0.4$, $\gamma_{\mathrm{d}} = -4$,  $A_{\mathrm{s}} = \SI{4}{\micro\kelvin\squared}$, $\alpha_{\mathrm{s}} = -0.9$,  $\beta_{\mathrm{s}} = -3$, $B_{\mathrm{s}} = 0.3$, $\gamma_{\mathrm{s}} = -4$, and $\rho_{\mathrm{ds}}~=~0.1$ that are picked to be approximately consistent with the posterior values shown in Fig.~\ref{fig:posteriors_pysm}. Note that the accuracy of the covariance matrix is not important for this test because the $\chi^2$ statistic is applied to both data and posterior predictive draws.} The test is performed by comparing
 $\chi^2(\bm{x}_{\mathrm{obs}})$  
to the distribution of $\chi^2$ values of the posterior predictive distribution:  $P(\chi^2(\bm{x}_{\mathrm{r}})  | \bm{x}_{\mathrm{obs}})$. Here, the observed data are the same as used for Fig.~\ref{fig:posteriors_pysm} and~\ref{fig:data_pysm}. For an accurate model, the $\chi^2(\bm{x}_{\mathrm{obs}})$ value should be consistent with a likely draw from the distribution.\footnote{The absolute values of $\chi^2$ are unimportant;  a unity reduced $\chi^2$ value is not expected because the data distribution is non-Gaussian, the fiducial parameter values used for the covariance matrix are off, and the covariance matrix estimate is noisy.} It can be seen that this is the case for all combinations of PySM models.

\section{Discussion}\label{sec:discussion}

The results from Sec.~\ref{sec:results_joint_nilc} show increased constraining power with the inclusion of additional sky component estimates for our joint NILC SBI setup. This appears to contradict the findings of  \cite{remazeilles_2021}, who demonstrate that deprojecting additional sky components increases the variance of the final $r$ estimate. The difference lies in the treatment of the NILC outputs: \cite{remazeilles_2021} only use the deprojected NILC CMB component map to perform $r$ inference. In our case (see Fig.~\ref{fig:sigma_r}) this choice also shows increased variance with additional deprojection. Our joint NILC method, however, keeps the foreground component map estimates. Interpreted as a data compression algorithm, the addition of components to the $\mathbf{A}$ matrix in Eq.~\eqref{eq:ilc} simply corresponds to a less aggressive compression. From this perspective, it should be expected that a larger number of sky components lowers the  uncertainty on $r$. Given our results, the inclusion of second-order variations in $\beta_{\mathrm{d}}$ and $\beta_{\mathrm{s}}$, as well as variations in the dust temperature, are natural extensions of the current setup. The only challenge would be the enlarged data vector, as discussed below.

The current setup requires roughly $5\cdot 10^{4}$ simulations, which is significantly above the usual  budget of $\mathcal{O}(10^3)$ for most CMB analyses.  Because simulation generation is not a limiting factor in our pipeline, we make no effort to reduce this number. Further compression of the 165-dimensional data vector offers the most straightforward path to lowering the simulation cost. Compression can be done linearly, using methods such as e-MOPED or Canonical Correlation Analysis \citep{park_2025}, or non-linearly using a neural network. In principle, the bias and weights $\bm{\psi}$ of such a compression network $g$ can be jointly optimized with the parameters $\bm{\lambda}$ of the normalizing flow by simply updating the loss function in Eq.~\eqref{eq:kl_divergence} from $-\sum_{i=1}^{N_{\mathrm{sim}}}\log q_{\bm{\lambda}}(\bm{\theta}_i | \bm{x}_i) / N_{\mathrm{sim}}$ to $-\sum_{i=1}^{N_{\mathrm{sim}}}\log q_{\bm{\lambda}}\left(\bm{\theta}_i | g_{\bm{\psi}}(\bm{x}_i)\right) / N_{\mathrm{sim}} $ \citep{barber_2003}. The latter approach has seen significant use in large-scale structure data analysis \citep{jeffrey_2021}. Future work should explore the viability of such compression schemes for the setup presented here.

Even without major reductions in the required number of simulations, the computational cost remain manageable if map-based methods can be used to generate noise and instrumental systematics simulations. Relatively sophisticated map-based noise simulations can be generated in $\mathcal{O}(1)$ CPU-seconds at these resolutions using the methods from \cite{atkins_2023}. The complex time-domain filtering can be forward modeled at similar cost using the observation-matrix approach from \cite{bk_obs_matrix_2016}. 
On the data compression side, the low map resolution should make Wiener-like filtering needed to undo the $E$- to $B$-mode mixing due to instrumental filtering, anisotropic noise, masking, and lensing cheap enough to stay within the computational budget. However, this will need to be demonstrated in future work.

Finally, the NPE estimator could be trained on a large number of cheap, low-fidelity map-domain simulations and then fine-tuned on a much smaller, $\mathcal{O}(10^3)$, set of expensive, high-fidelity simulations. This would be relevant, for instance, when certain crucial instrumental systematics can only be modeled using expensive time-domain simulations. Such pretraining, where the weights and biases are simply initialized from a network trained on the low-fidelity simulations, has been shown to provide significant, $\mathcal{O}(10)$, reductions in the number of simulations \citep{saoulis_2025}. Further improvements might be possible with more sophisticated  approaches \citep{thiele_2025}. Given the effectiveness of cheap map-domain simulations for CMB analyses, it would be interesting to apply this approach to our setup.  

We have investigated the potential of sequential NPE (SNPE) to reduce the number of simulations. SNPE iteratively transforms the proposal distribution used to draw the simulations from the prior to the posterior. The potential advantage is a more efficient use of simulations. The downside is a loss of amortization: the final NPE estimator $q$ can only be conditioned on the observed data, making coverage tests computationally difficult. We describe the SNPE setup and results in more detail in Appendix~\ref{sec:snpe}. In summary, we  only obtain modest gains from the use of SNPE for our problem. We find some differences when comparing the NPE and SNPE posteriors. Together with the coverage tests for the marginal distributions of the foreground parameters in Appendix~\ref{app:coverage}, this indicates that the NPE has not fully converged. 
This manifests itself mainly as a slight suboptimality in the synchrotron parameters and does not meaningfully affect the cosmological results. 
We suspect that this is due to the relatively large data vector size. We conclude that SNPE has limited use for our setup and that compressing the data further should be a priority for future work. 

The NPE approach based on normalizing flows does not scale to large parameter spaces. With the same architecture, it will likely be difficult to significantly extend beyond the current 13-dimensional case. For realistic analyses, with a relatively large number of nuisance parameters describing instrumental systematics, this might appear to be a problem. 
However, SBI does not require the number of parameters included in the NPE to match the number of nuisance parameters. There may be significantly more nuisance parameters; SBI will automatically marginalize over them. Because this feature might seem surprising, Appendix~\ref{app:nuisance} provides a derivation and an example showing that we obtain consistent posteriors for $r$ and $A_{\mathrm{lens}}$ regardless of whether we train the NPE on all 13 parameters of the forward model, or just on the $r$ and $A_{\mathrm{lens}}$ parameters. This demonstrates that the method is, in fact, able to incorporate large numbers of nuisance parameters, actually in an easier way than likelihood-based methods. For future work, it would be interesting to extend the compressed data vector with components that are designed to be sensitive to particular nuisance parameters, such as the inclusion of differences between detector arrays to become sensitive to certain array-specific systematics. 

Beyond the addition of more instrumental realism to the pipeline, future work could also focus on alternative, more realistic, foreground models. Given the posterior predictive checks in Sec.~\ref{sec:pysm_results}, it would be interesting to explore the addition of spatial correlations between the amplitude and spectral index fields. It would be good to check for biases due to polarized anomalous microwave emission \citep{draine_1998} and  CO line emission \citep{goldreich_1981,puglisi_2017}, and explore their inclusions in the forward model.
In addition, the results in Sec.~\ref{sec:results_joint_nilc} motivate the inclusion of more realistic foreground non-Gaussianity and anisotropy. Given that the joint NILC compression lowers the uncertainty on $r$ by 20\% compared to the purely harmonic joint HILC compression when the non-Gaussianity due to spectral index anisotropy is large (see Fig.~\ref{fig:sigma_r}), one would expect further relative improvements in the uncertainty on $r$ once more realistic foreground non-Gaussianity and anisotropy are incorporated in the forward model. This expectation is consistent with \cite{surrao_2024}, who found that, with simulations consisting of the CMB and a bright, highly non-Gaussian Compton-$y$ component, SBI with joint NILC compression significantly improves the statistical power compared to purely harmonic-based compression methods. The absence of an improvement in $\sigma_r$ when comparing NILC to HILC for the PySM test data in Sec.~\ref{sec:pysm_results} is likely a consequence of the NPE being trained on simulations that are statistically isotropic and only weakly non-Gaussian. 

To move towards more realistic foreground non-Gaussianity, it would be interesting to include the dust filament model by \cite{hervias_caimapo_2022} in the simulations. 
In terms of including prior information on the spatial distribution of the Galactic emission, an appealing approach would be to use a version of the $\texttt{d12}$ PySM model \citep{martinez_2018} or the dust model from \citep{odea_2011}, potentially updated to use the more recent 3D dust reconstruction from \cite{edenhofer_2024}. Such an approach would effectively update the isotropic prior on the dust amplitude that we currently use to an anisotropic one based on external data. Unlike the existing $\texttt{d12}$ model, whose hyperparameters are fixed, one would use the data to constrain them. The ability of our SBI framework to easily include such external data as a prior should naturally extend to future 3D reconstructions of the Galactic magnetic field from, for example, the starlight polarization survey PASIPHAE  \citep{tassis_2018}.

\section{Conclusion} 

This exploratory work on $r$ inference using the NILC foreground cleaning method integrated into an SBI pipeline reveals three main points, as summarized in the following.

(i) When trained on foreground simulations drawn from the relatively simple foreground model from \cite{azzoni_2021}, which parameterizes the spatial variations of the dust and synchrotron spectral indices as two isotropic Gaussian random fields with power-law power spectra, the method produces unbiased $r$ inference for all considered PySM  \citep{pan_exp_2025} models, including the high-complexity $\texttt{d12}$ model. We explain and demonstrate that this is made possible due to  the NILC-based compression of the data that is designed to suppress the effect of mismodeling in the simulations, and the ability of SBI to perform inference with this compressed data. The inference is done for a five-year SO-like survey using an enlarged, $f_{\mathrm{sky}} = 0.21$, sky mask
with the goal/optimistic noise levels of the SO SAT specification~\citep{so_forecast_2019}.\footnote{Since we do not increase the SO noise levels to account for our sky fraction, which is approximately twice that used in the SO forecasts, our achieved uncertainties on $r$ should not be directly compared to those in \cite{wolz_2024}.} 
For this setup, we show that the standard multi-frequency power-spectrum-based likelihood method from e.g.~\cite{bicepkeck_2021} is unusable, already exhibiting significant bias when the mild foreground SED anisotropies suggested by the low-complexity PySM models are present. More importantly, for a similar instrumental setup with comparable per-pixel noise level but a sky coverage of only 10\%, \cite{wolz_2024} shows that the standard NILC method exhibits an $8.3\sigma$ bias when applied to the medium-complexity $\texttt{d10\_s5}$ PySM simulation. For these foregrounds, the generalized power-spectrum-based $C_{\ell}$-moments method and the map-based parametric method with additional marginalization over dust contamination still achieve approximately unbiased results, $1.0\sigma$ and $0.3\sigma$, respectively. 
The high-complexity $\texttt{d12}$ model, not yet considered in the SO analysis context,
poses a more significant challenge for existing pipelines \citep{carones_2023,bianchini_2025}.  
As a final comparison, \cite{liu_2025} find a 3$\sigma$ bias for the artificially complex $\texttt{d10x1\!.\!6\_s5}$ model for the $C_{\ell}$-moments method, which our method reduces to $1\sigma$ despite using observations with significantly lower instrumental noise. Our results demonstrate, for the first time, that a single analysis method achieves unbiased $r$ results across all considered PySM foreground models, irrespective of complexity and without foreground-model–specific tuning, for ground-based, wide-area observations like those considered here. 

(ii) The simultaneous use of CMB and foreground component ILC estimates significantly improves the constraining power on $r$ compared to the NILC and constrained NILC methods that only use the foreground-cleaned CMB estimate. Our proposed SBI method, which uses NILC estimates of CMB, dust, and synchrotron amplitudes, and their (linearized) spectral indices, consistently outperforms the NILC and constrained NILC methods. Depending on the foreground complexity and deprojection choices, the  standard deviation of the $r$ posterior is reduced by a factor 1.2-1.5 compared to NILC and by 2.3-4.1 compared to the constrained NILC method.  

(iii) Map-based foreground cleaning appears important for achieving unbiased $r$ inference with wide-area ground-based observations, unless foreground complexity is at the minimum allowed level. We demonstrate that biases appear once we replace the map-based NILC compression with the purely harmonic HILC compression in our SBI pipeline.  
Given the equivalence between the HILC and multi-frequency power-spectrum approaches \citep{surrao_2024}, these results confirm that even an extended multi-frequency power-spectrum-based approach, one that includes the accurate marginalization over foreground uncertainty that our SBI method enables, will face limitations as a  robust inference method for ground-based wide-area observations.

In summary, our proposed method appears to solve the problem of component separation for $r$ inference with an observatory similar to the SO SAT configuration, even in the presence of highly complex foregrounds. 
Since we test the method on a 1.5-2 times larger sky region ($f_{\mathrm{sky}} = 0.21$) than typically considered by SO, our unbiased results strongly support scalability to the nominal SO case. 
We demonstrate the potential and feasibility of an entire map-to-parameters $r$ inference pipeline based on SBI. While our current setup is  idealized in terms of instrumental systematics, we expect that, as long as these effects can be included in the simulations, they will not spoil our ability to obtain unbiased results. In fact, most instrumental complexities are likely to reduce the impact of foreground mismodeling by increasing the systematic error budget. 
Still, this expectation will have to be validated by including more instrumental realism in future work. The SBI framework is well-suited for this purpose, as it eliminates the need to construct and sample over simplified, low-dimensional parameterizations of systematics. Instead, it allows any systematic to be properly marginalized over as long as it can be simulated. In addition, the presented framework provides a natural way to explore more realistic and complex foreground models that are difficult to incorporate into traditional likelihood-based approaches.

\begin{acknowledgments}

We thank William R.\ Coulton, David N.\ Spergel, Fiona McCarthy, Noemi Anau Montel, Eiichiro Komatsu, Patricia Diego Palazuelos, Florie Carralot, and Martin Reinecke for helpful discussions and Jo Dunkley, 
Josquin Errard, and the Simons Observatory Theory and Analysis Committee for providing comments on the manuscript. 
JCH acknowledges support from DOE grant DE-SC0011941, NASA grants 80NSSC22K0721 [ATP] and 80NSSC23K0463 [ADAP], and the Sloan Foundation.
This work has also received funding from the European Union's Horizon 2020 research and innovation programme under the Marie Skłodowska-Curie grant agreement no.\ 101007633. 
Computations were performed on the HPC systems Rusty and Raven at the Flatiron Institute and the Max Planck Computing and Data Facility, respectively. Some of the results in this paper have been derived using the healpy and HEALPix packages. This work is not an official
Simons Observatory paper.

\end{acknowledgments}

\software{
BlackJax \citep{cabezas_2024},
GetDist \citep{lewis_2019},
DUCC,\footnote{\url{https://gitlab.mpcdf.mpg.de/mtr/ducc}}
HEALPix \citep{healpix_2005},
healpy \citep{zonca_2019},
JAX \citep{jax2018github},
matplotlib \citep{hunter_2007},
mpi4py \citep{dalcin_2005},
numpy \citep{harris_2020},
Optuna \citep{akiba_2019},
pixell \citep{naess_2021},
pyilc \citep{mccarthy_2024},
PySM \citep{thorne_2017,zonca_2021,pan_exp_2025},
PyTorch \citep{pytorch_2019},
sbi \citep{sbi_2025},
scipy \citep{scipy_2020}, 
tarp \citep{lemos_2023}
}

\appendix

\section{Instrumental setup}\label{app:instrument}

The noise power spectra that describe the statistics of the noise $\bm{n}$ in Eq.~\eqref{eq:cmb_data_model} are modeled as follows:
\begin{align}\label{eq:noise_nl}
    N_{\ell} = \frac{\pi^2\sigma_{\nu}^2 }{(60\cdot 180)^2}\left[ 1 + \left( \frac{\ell}{\ell_{\mathrm{knee}}^{\nu}}\right)^{\alpha_{\nu}}\right ]\, ,
\end{align}
where $\sigma_{\nu}$ is the white noise standard deviation in $\SI{}{\micro\kelvin} \, \mathrm{arcmin}$. The $\ell_{\mathrm{knee}}$ and $\alpha$ parameters describe how the correlated noise power increases towards low $\ell$ values. For the SO-like maps we use the prescription from \cite{wolz_2024} with the ``goal'' and ``optimistic'' values for $\sigma_{\nu}$ and $\ell_{\mathrm{knee}}$, respectively. For the \emph{WMAP} and \emph{Planck} bands we only consider the white noise component in Eq.~\eqref{eq:noise_nl}, which is a good approximation for the intermediate scales ($30 \leq \ell \leq 200$) that we use throughout this work. The \emph{Planck} noise levels are taken from Table~4 in \cite{planck_2018_I} and the \emph{WMAP} noise levels are measured from the average power spectrum (for $30 \leq \ell \leq 300$) of noise simulations created by scaling the hit maps from the DR5 \emph{WMAP} release\footnote{\url{https://lambda.gsfc.nasa.gov/product/wmap/dr5/}} with the ``Uncleaned $\sigma_0(Q, U )$'' values from Table~5 from \cite{bennett_2013}.  See a summary of the noise parameters in Table~\ref{table:instr_params}.

The instrumental beams,  $\mathbf{B}$ in Eq.~\eqref{eq:cmb_data_model}, are Gaussian specified by full width at half maximum (FWHM) parameters taken from \cite{so_forecast_2019, planck_2018_I} and the DR5 \emph{WMAP} LAMBDA page.\footnote{We mistakenly interpreted the \emph{WMAP} ``Beam Size'' values as FWHM instead of the square root of the beam solid angle. For a Gaussian beam this overestimates the beam size by $\sim 6\%$.}
The beam parameters are summarized in Table~\ref{table:instr_params}.

\begin{deluxetable}{c|cccc}
\tablecaption{Overview of noise and beam parameters for each frequency band.\label{table:instr_params}}
\tablehead{$\nu$  [\SI{}{\giga\hertz}]& \colhead{$\sigma$ [\SI{}{\micro\kelvin} $\mathrm{arcmin}$]} & \colhead{$\alpha$ } & $\ell_{\mathrm{knee}}$ & FWHM [$\mathrm{arcmin}$]
}
\startdata
     23 & 290.6 & - & - & 52.8\\
     27 & 33.0 & -2.4 & 15 & 91.0\\
     39 & 22.0 & -2.4 & 15& 63.0\\
     93 & 2.5 & -2.5 & 25& 30.0\\
     145 & 2.8 & -3.0 & 25& 17.0\\
     225 & 5.5 & -3.0 & 35& 11.0\\
     280 & 14.0 & -3.0 & 40& 9.0\\
     353 & 365.5 & - & - & 4.92\\
\enddata
\end{deluxetable}

The high-pass filter $\mathbf{F}$ in Eq.~\eqref{eq:cmb_data_model} smoothly transitions from zero to one between $\ell_{\mathrm{min}} - \Delta\ell$ and $\ell_{\mathrm{min}}$. It is implemented as a diagonal matrix in the harmonic domain with diagonal given by:
\begin{align}
\begin{split}
    f_{\ell} =& \frac{1}{2} \left[1 - \cos\left(\frac{\pi\ell}{\Delta\ell} \right) \right] \mathbbm{1}(\ell_{\mathrm{min}} -\Delta\ell< \ell < \ell_{\mathrm{min}}) \\ &+\mathbbm{1}(\ell \geq \ell_{\mathrm{min}})\, ,
\end{split}
\end{align}
with $\ell_{\mathrm{min}} = 30$ and $\Delta\ell = 5$. The main purpose of the filter is to make sure that signal at multipoles below $\ell_{\mathrm{min}}$ does not contribute to the NILC covariance matrix, mimicking the filtering operations typically performed by ground based CMB observatories.

\begin{figure}[t]
    \centering \includegraphics[width=\columnwidth]{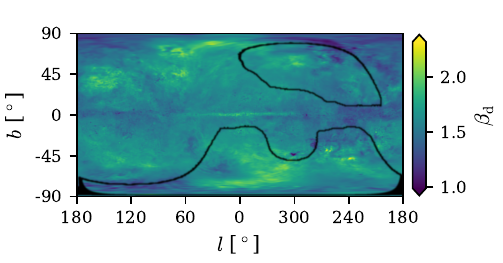}       
    \caption{The edge of the mask used in Sec.~\ref{sec:pysm_results} overlaid on top of the dust spectral index ($\beta_{\mathrm{d}}$) map of the PySM \texttt{d10} model in Galactic coordinates. The black line denotes the outer edges of the mask. The mask is apodized inwards by \SI{10}{\degree} using a cosine profile.}\label{fig:mask}
\end{figure}

The mask matrix $\mathbf{M}$ in Eq.~\eqref{eq:cmb_data_model} is simply the identity matrix except when the setup is tested on the PySM simulations in Sec~\ref{sec:pysm_results}. The mask used there is the combination of an enlarged version of the SO mask from \cite{so_forecast_2019} and the (no-apodization) Galactic plane mask provided by the \emph{Planck} collaboration that leaves 70\% of the sky unmasked.\footnote{\texttt{HFI\_Mask\_GalPlane-apo0\_2048\_R2.00.fits} from the Planck Legacy Archive: \url{https://pla.esac.esa.int}.} We apodize the combined mask inwards by \SI{10}{\degree} using a cosine profile. The fraction of sky weighted by the mask $m(\hat{n})$, $\int \mathrm{d}\hat{n} \, m(\hat{n}) / (4 \pi)$, is approximately 0.21. The mask is overlaid on the PySM \texttt{d10} dust spectral index map in Fig.~\ref{fig:mask}.

\section{Notation for Spherical Harmonics}\label{app:sht}

Throughout this work, we distinguish between~$\mathbf{Y}$, which denotes a spin-0 spherical harmonic transform and $\widetilde{\mathbf{Y}}$, which denotes the spin-weighted transformation from $E$- and $B$-mode spherical harmonic coefficients to  Stokes $Q$ and $U$ fields. 

The spin-0 transform (synthesis) of a set of $N_{\mathrm{harm}} = (\ell_{\mathrm{max} }+1)^2$ spherical harmonic coefficients $\bm{x}$ is represented by:
\begin{align}
(\mathbf{Y}\bm{x})_p  = \sum_{\ell=0}^{\ell_{\mathrm{max}}} \sum_{ m=-\ell}^{\ell}  Y_{p, \ell m} \, x_{\ell m}  \, ,
\end{align}
where $Y_{p, \ell m}$ are the spherical harmonics evaluated at the spherical coordinates of pixel $p$.  Spherical harmonic analysis applied to a pixelized input map $\bm{x}_{\mathrm{pix}}$ is then described by:
\begin{align}
\left(\mathbf{Y}^{\dagger} \mathbf{W}\bm{x}_{\mathrm{pix}}\right)_{\ell m} = \sum_{p=1}^{N_{\mathrm{pix}}} w_p\,  x_p Y^*_{p, \ell m} \, ,
\end{align}
where $N_{\mathrm{pix}}$ denotes the number of pixels used to discretize the sphere. The diagonal matrix of integration weights $\mathbf{W}$ is such that $\mathbf{Y}^{\dagger} \mathbf{W} \mathbf{Y} = \mathbf{1}$. 

The $\widetilde{\mathbf{Y}}$ operation should be understood as transforming from the $E$- and $B$-mode spherical harmonic coefficients \citep{zaldarriaga_1997,kamionkowski_1997} to the $Q$ and $U$ Stokes parameters in pixel space. In this case $\widetilde{\mathbf{Y}}$ is a $(2 \times N_\mathrm{pix}) \times (2 \times N_{\mathrm{harm}})$ complex matrix and the $\widetilde{\mathbf{Y}}\bm{x}$ operation is given by the block matrix multiplication: 
\begin{align}
\begin{pmatrix} Q_p \\ U_p \end{pmatrix} = \sum_{\ell=0}^{\ell_{\mathrm{max}}} \sum_{ m=-\ell}^{\ell} \! \! \begin{pmatrix}  -X_{p, \ell m} & \mathrm{i} Z_{p, \ell m} \\  -\mathrm{i} Z_{p, \ell m} & - X_{p, \ell m} \end{pmatrix} \! \! \begin{pmatrix}   x_{\ell m}^E \\ x_{\ell m}^B \end{pmatrix} \, ,
\end{align}
where we defined the following linear combinations of spin-weighed spherical harmonics \citep{newman_1966,goldberg_1967}:
\begin{align}
X_{p, \ell m} &= \frac{1}{2} \left( {}_2 Y_{p,\ell m} + {}_{-2} Y_{p,\ell m} \right) \, , \\
Z_{p, \ell m} &= \frac{1}{2} \left( {}_2 Y_{p,\ell m} - {}_{-2} Y_{p,\ell m} \right) \, .
\end{align}
The $\widetilde{\mathbf{Y}}^{\dagger} \mathbf{W} \bm{x}_{\mathrm{pix}}$ operation is then a shorthand for:
\begin{align}
\begin{pmatrix} x_{\ell m}^E \\ x_{\ell m}^B \end{pmatrix} = \sum_{p=1}^{N_{\mathrm{pix}}} w_p \! \! \begin{pmatrix}  -X^*_{p, \ell m} & \mathrm{i} Z^*_{p, \ell m} \\ -\mathrm{i} Z^*_{p, \ell m} & - X^*_{p, \ell m} \end{pmatrix} \! \! \begin{pmatrix}  Q_p \\ U_p \end{pmatrix} \, .
\end{align}
The $\widetilde{\mathbf{Y}}^{\dagger} \mathbf{W} \widetilde{\mathbf{Y}} = \mathbf{1}$ 
identity also holds.

\section{Hyperparameter optimization}\label{app:hyperparam_tuning}

To avoid overly conservative posterior estimates, we find that it is important to optimize certain hyperparameters of the MAF architecture \citep{papamakarios_2017} and the Adam optimizer \citep{kingma_2014}.  For every NPE setup in this paper, we therefore perform an automated hyperparameter search by training 144 versions of the NPE with unique hyperparameters and selecting the setup with the best loss on a held-out validation set comprised of 10\% of the total parameter-data pairs. The search is performed using the multivariate tree-structured Parzen Estimator  algorithm \citep{watanabe_2023} implemented in the Optuna package \citep{akiba_2019}. The hyperparameters are the constant base learning rate of the Adam optimizer: $\gamma \in [10^{-6}, 10^{-2}]$; the number of transformations of the normalizing flow, i.e. the number of ``Masked Autoencoder for Distribution Estimation'' (MADE) \citep{germain_2015} autoregressive blocks, $N_{\mathrm{MADE}}\in [3, 13]$; the number of layers of each (simple non-residual) MADE block: $N_{\mathrm{layer}}\in [1, 7]$; and the number of nodes in each hidden layer of the MADE networks: $N_{\mathrm{hidden}}\in [10, 100]$. Fixed hyperparameters include the batch size, which is set to 256, and the $\beta_1$, $\beta_2$, and $\epsilon$ parameters of the Adam optimizer, which are kept to their defaults of 0.9, 0.999 and $10^{-8}$. The MADE network uses hyperbolic tangent activation. We do not use dropout or batch normalization during training, but do clip the total gradient above \SI{6.0}{}{} as mild regularization. To avoid overfitting, we employ an early stopping criterion that halts training when the validation loss has not decreased in the last 30 epochs, with a maximum number of epochs set to \SI{1000}{}{}.

\section{Additional coverage tests}\label{app:coverage}

\begin{figure}[htpb]
\includegraphics[width=\columnwidth]{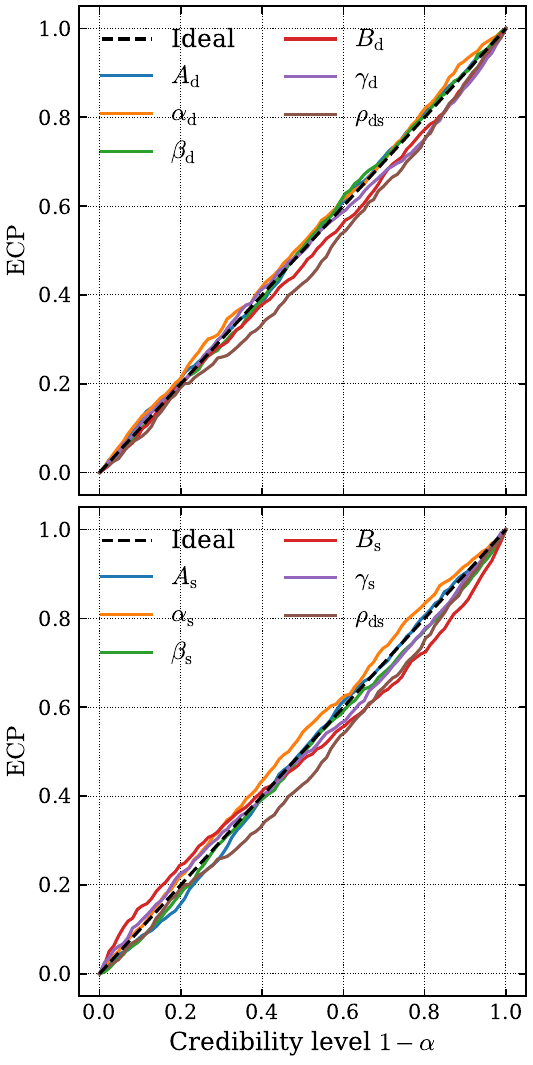}       
    \caption{TARP coverage tests. Top: the TARP test applied to the marginal posteriors of the dust parameters. Bottom: the test applied to the marginal posteriors of the synchrotron parameters. }\label{fig:tarp_marg_fg}
\end{figure} 

\begin{figure}[htpb]
\includegraphics[width=\columnwidth]{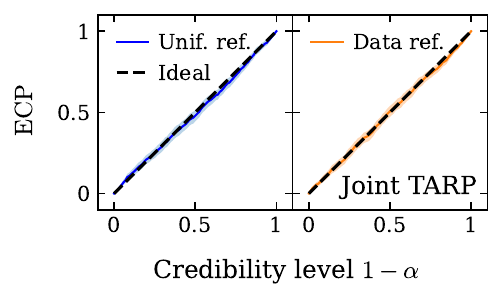}
    \caption{Comparison of the joint coverage determined using a uniform TARP reference distribution that is independent of the data (left) and one that is dependent on the data (right). }\label{fig:tarp_emoped}
\end{figure} 

In this appendix, Fig.~\ref{fig:tarp_marg_fg} shows additional coverage tests for the foreground parameters and Fig.~\ref{fig:tarp_emoped} provides a comparison between the use of two TARP reference distributions.

\begin{figure*}[t]
    \centering \includegraphics[width=\textwidth]{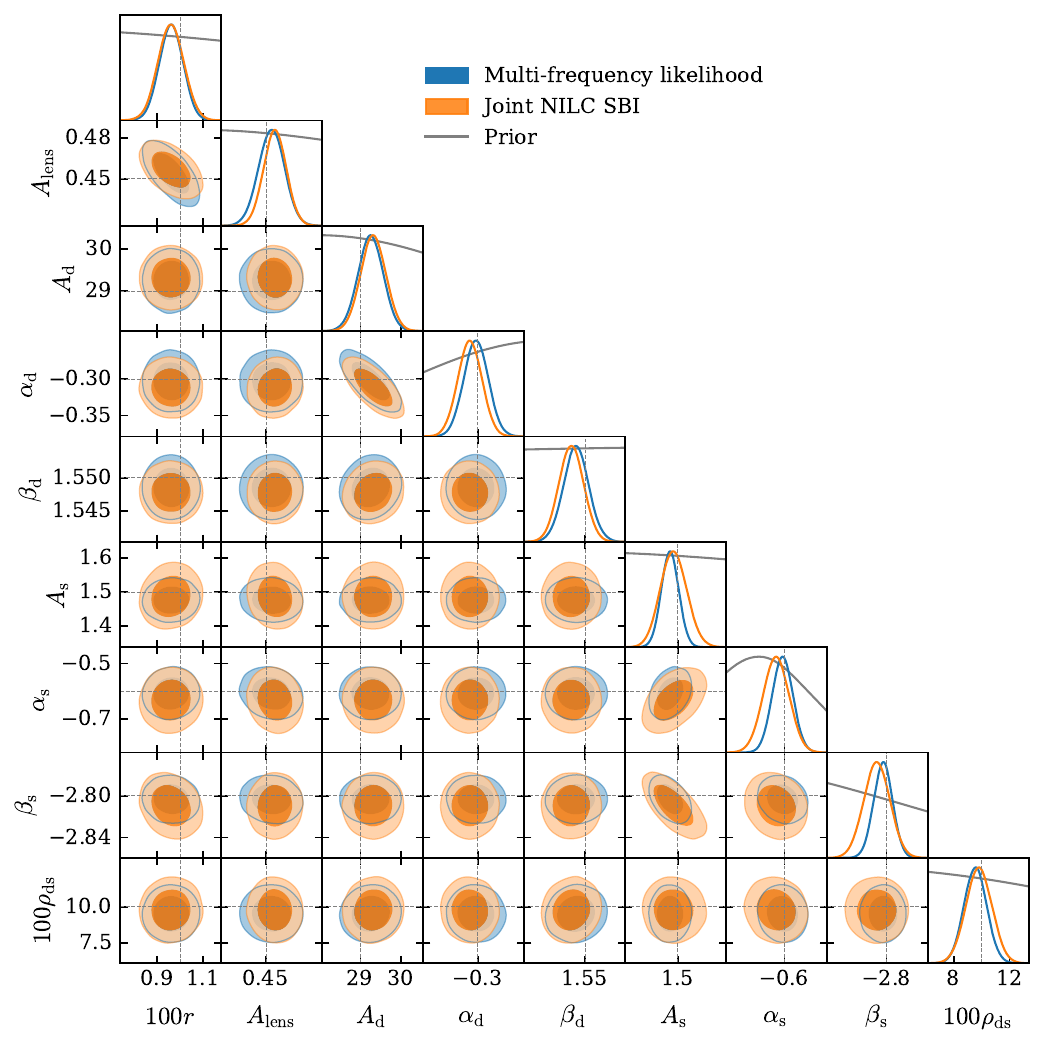}       
    \caption{Comparison between the posterior distributions obtained with the multi-frequency power-spectrum based likelihood approach and our SBI method. Both use the same data, generated without spatial variation in the spectral index fields or other source of statistical anisotropy, making the likelihood approach statistically optimal. This is the nine-parameter version of Fig.~\ref{fig:mcmc_cosmo_only}.}\label{fig:fig:mcmc_full}
\end{figure*}

\section{Multi-frequency power spectrum likelihood}\label{app:mflike}

We describe the analytical prescription for the likelihood and covariance matrix of the multi-frequency power spectrum likelihood. Similar to real analyses, we approximate the likelihood as Gaussian with a known covariance matrix that is evaluated at a fiducial parameter vector.

The Gaussian likelihood for the binned \emph{BB} power spectra is as follows:
\begin{align}\label{eq:ana_like}
\begin{split}
    &-2\log  P(\hat{C}_{b} | \bm{\theta}) \propto \sum_{b, b'}\sum_{\alpha, \alpha'}\\ &\quad \ \left[\hat{C}^{\alpha}_{b} - C^{\alpha}_{b}(\bm{\theta})\right]^{\top}\! \left(\Sigma^{\alpha \alpha'}_{bb'}\right)^{-1}\left[\hat{C}^{\alpha'}_{b'} - C^{\alpha'}_{b'}(\bm{\theta})\right] \, ,
\end{split}
\end{align}
where $\hat{C}^{\alpha}_{b}$ and $C^{\alpha}_{b}(\bm{\theta})$ are respectively the measured and theoretical binned power spectrum for a bin index $b$ and frequency-split combination $\alpha = \{\nu_1, \nu_2, i, j \}$. Here, $\nu_1$ and $\nu_2$ refer to the two frequency bands that are correlated and $i$ and $j$ refer to the two noise splits that are correlated. The measured spectra are created following the procedure in Sec.~\ref{sec:sim_setup} but without the NILC compression step. The cross-spectra are thus between frequency maps instead of sky components. 

The covariance matrix in Eq.~\eqref{eq:ana_like} is computed as follows:
\begin{align}
\Sigma^{\alpha \alpha'}_{bb'} = \frac{1}{\Delta \ell(b)}  (\Phi^{\alpha \beta}\Xi^{\beta \beta'}_{b}\Phi^{\beta' \alpha'}) \delta_{bb'} \, ,
\end{align}
where $\Delta \ell(b) = \sum_{\ell \in b} (2 \ell + 1)$. The $\Xi^{\beta \beta'}$ matrix for given $\beta = \{\nu_1, \nu_2, i, j \}$ and $\beta' = \{\nu'_1, \nu'_2, i', j'\}$ is given by:
\begin{align}
\begin{split}
    \Xi^{\beta \beta'} = \
    &C_b^{\nu_1 \nu'_1} \delta_{i i'}  C_b^{\nu_2 \nu'_2} \delta_{j j'} \left[(1 - \delta_{i j})(1-\delta_{i' j'}) \right] \\
    + \ &C_b^{\nu_1 \nu'_2} \delta_{i j'} C_b^{\nu_2 \nu'_1} \delta_{j i'} \left[(1 - \delta_{i j})(1-\delta_{i' j'} )\right] \, .
\end{split}
\end{align}
The $\Phi^{\alpha \beta}$ matrices average spectra with $\nu_1, \nu_2$ together with spectra with $\nu_2, \nu_1$. This is a consequence of the co-added spectra we use, see Eq.~\eqref{eq:cross_spectra}. 
We neglect the parameter dependence of the covariance matrix and instead evaluate it for a fiducial parameter vector $\bm{\theta}^{\mathrm{fid}}$, which we, for simplicity, set to the true parameters. This approach yields slightly overly optimistic likelihood results compared to a real analysis that would use an estimate of the parameters, but we expect this to be a relatively minor effect. 

We implement the likelihood in Eq.~\eqref{eq:ana_like} in JAX \citep{jax2018github} and use Hamiltonian Monte-Carlo (HMC) implemented in BlackJAX \citep{cabezas_2024} to sample the posterior. We use the automated HMC hyperparameter tuning as implemented in BlackJAX, which has been adapted from the Stan package \citep{stan_software}. We initialize ten chains from draws from the prior and run the chains for \SI{4000}{}{} steps each after a burn-in and tuning phase. The setup results in Gelman-Rubin $R-1$ statistics \citep{gelman_1992} that are $<0.01$ for all parameters, indicating that the chains are converged.

\section{Comparison with the multi-frequency likelihood}\label{app:full_mf_comparison}

In Fig.~\ref{fig:fig:mcmc_full} we show the full nine-parameter version of the comparison shown in Fig.~\ref{fig:mcmc_cosmo_only}. There is good agreement for all parameters. The marginal posteriors on the $A_{\mathrm{s}}$, $\alpha_{\mathrm{s}}$, and $\beta_{\mathrm{s}}$ are slightly wider for the NPE method. This is likely caused by a suboptimal  fiducial synchrotron spectral index parameter in the NILC compression. 

\section{Inferred dust complexity for the PySM test}\label{app:pysm_sbd}

\begin{figure}[t]
\includegraphics[width=\columnwidth]{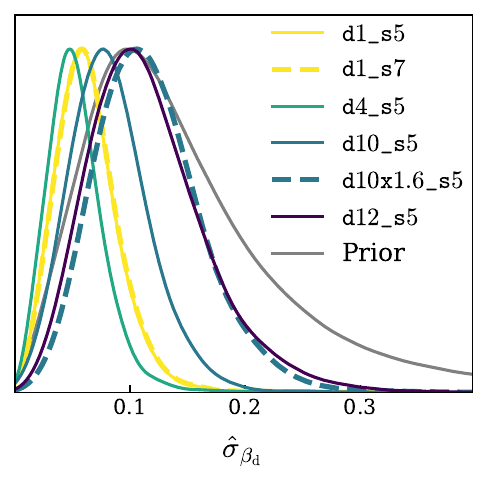}       
    \caption{Inferred standard deviation of the dust spectral index field for each of the PySM test cases shown in Fig.~\ref{fig:posteriors_pysm}.
    }\label{fig:posteriors_pysm_sbd}
\end{figure} 

Fig.~\ref{fig:posteriors_pysm_sbd} shows the estimated standard deviation of the dust spectral index field for each of the PySM test cases shown in Fig.~\ref{fig:posteriors_pysm}. They are created by converting the $B_{\mathrm{d}}$ and $\gamma_{\mathrm{d}}$ posterior samples to $\sigma_{\beta_{\mathrm{d}}}$ using Eq.~\eqref{eq:beta_var}. The transformation reveals that, while the $B_{\mathrm{d}}$ posteriors in Fig.~\ref{fig:posteriors_pysm} do not significantly depart from zero, the standard deviation is clearly detected. The observed hierarchy between the models is the same as mentioned in Sec.~\ref{sec:pysm_results}. It is also revealed how the prior is relatively informative in this transformed space. In future work, less restrictive priors could be used to avoid biasing the inference towards the prior mean. 

\section{Experiment with Sequential NPE}\label{sec:snpe}

\begin{figure*}[htbp]
    \centering \includegraphics[width=\textwidth]{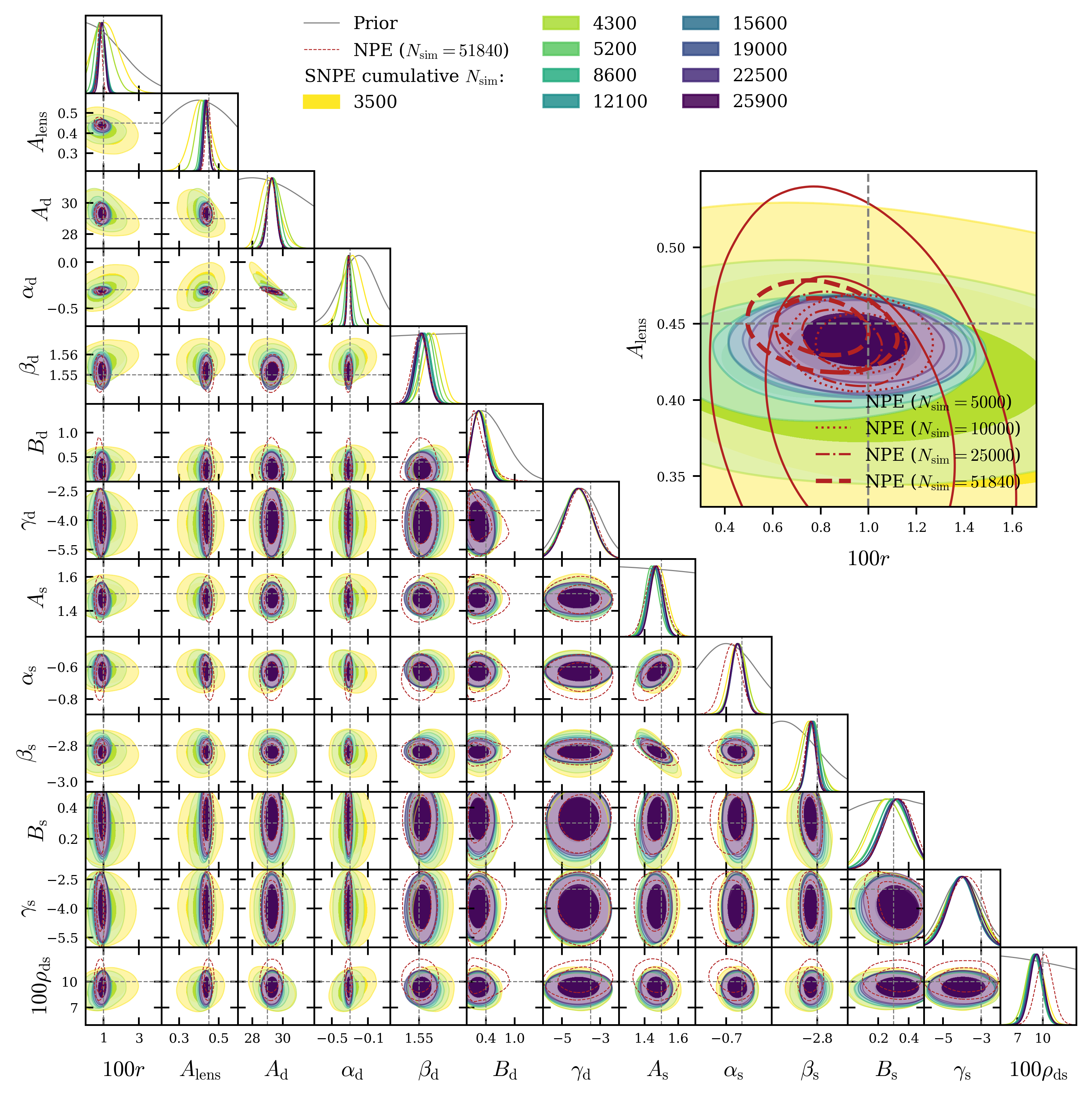}       
    \caption{Comparison between the posterior distributions obtained with Neural Posterior Estimation (NPE) (dashed red lines) and Sequential NPE (SNPE). The colors indicate the different rounds of the SNPE method. The first three rounds have been omitted. The insert shows a zoomed-in version of the marginal distribution of $r$ and $A_{\mathrm{lens}}$ with three extra red curves that show NPE with fewer simulations.}\label{fig:snpe}
\end{figure*} 

We explore the potential of sequential NPE (SNPE) to reduce the number of simulations. SNPE is an iterative algorithm that starts by fitting $q(\theta | \bm{x})$ using the regular NPE algorithm applied to a relatively small number of $\bm{\theta}$-$\bm{x}$ pairs with $\bm{\theta}$ drawn from the prior. The resulting posterior estimate $q(\bm{\theta} | \bm{x}_{\mathrm{obs}})$, which is computed for the actual observed data, is then used as the proposal distribution for a subsequent round of NPE (instead of the prior). This procedure is typically repeated for several rounds, with every new round using the posterior from the previous round as the proposal.\footnote{Depending on the SNPE implementation, a correction is applied to the proposals for the subsequent rounds to correct for the fact that the proposal is not equal to the prior anymore, see e.g. \cite{greenberg_2019}.} As the posterior estimates improve, the simulated data are drawn from $\bm{\theta}$ that lie closer to the posterior in parameter space and $q(\bm{\theta} | \bm{x}_{\mathrm{obs}})$ converges to the true posterior. The idea is that training SNPE requires
fewer simulations as $q$ does not have to generalize to irrelevant data far away from the true data. This advantage is simultaneously a disadvantage: the posterior estimator is not ``amortized'', it only applies to the single observed data vector used during training, making coverage tests computationally infeasible.

For our tests we use the Truncated SNPE (TSNPE) algorithm from \cite{deistler_2022} with a truncation level of $\epsilon=10^{-4}$. We run six rounds with 864 simulations each, followed by six rounds of 3456 simulations. The simulations from earlier rounds are re-used during training. The results are displayed in Fig.~\ref{fig:snpe}. Generally, we find good agreement between NPE and SNPE, although the $r$ posterior is shifted by approximately one standard deviation. The SNPE posterior becomes comparable to the NPE results after approximately $\SI{10000}{}{}$ simulations for most parameters and even provides slightly tighter constraints for some parameters ($B_{\mathrm{d}}$, $A_{\mathrm{s}}$ and $\alpha_{\mathrm{s}}$), suggesting that the NPE method has not fully converged. The marginal distribution for the $r$ parameter seems to converge at an increased rate for the SNPE method during the initial \SI{5000}{}{} simulations, already reaching a posterior width that is only twice the final width. However, convergence seems to slow down and proceeds at a similar rate for the SNPE and NPE methods for $N_{\mathrm{sim}} > \SI{5000}{}{}$. 

\section{Nuisance parameters}\label{app:nuisance}

Within the SBI framework nuisance parameters are automatically marginalized over when they are included in the simulations. This is in contrast with likelihood-based methods where nuisance  parameters have to be explicitly integrated over, e.g.\ included in the MCMC sampler. In this appendix we provide a simple derivation and demonstrate that this holds for our setup.

Consider the situation where $\theta_1$ represents a parameter of interest and $\theta_2$ is a nuisance parameter. We optimize the hyperparameters $\bm{\lambda}$ of a conditional normalizing flow $q_{\bm{\lambda}}(\theta_1 | \bm{d})$ such that the forward KL divergence between the marginalized distribution: $P(\theta_1|\bm{d})P(\bm{d}) = \int \mathrm{d}\theta_2 P(\theta_1, \theta_2 | \bm{d}) P(\bm{d})$ and $q_{\bm{\lambda}}(\theta_1 | \bm{d}) P(\bm{d})$ is minimized. It is straightforward to show that the optimal $\bm{\lambda}$ parameters are those that maximize $\log q_{\bm{\lambda}}(\theta_{1,i}|\bm{d}_i)$ when averaged over joint draws $\{(\bm{d}_{i}, \theta_{1,i})\}^{N_{\mathrm{sim}}}_{i=1}$ from the $P(\theta_1, \theta_2, \bm{d})$ distribution:  
\begin{align}
\begin{split}
    &\argmin_{\bm{\lambda}} \, D_{\mathrm{KL}}  \biggl( P(\theta_1 | \bm{d}) P(\bm{d}) \, || \, q_{\bm{\lambda}}(\theta_1 | \bm{d}) P(\bm{d}) \biggr) \\
    &\ =  \argmin_{\bm{\lambda}} \, \int P(\theta_1 | \bm{d}) P(\bm{d}) \log \frac{P(\theta_1|\bm{d})}{q_{\bm{\lambda}}(\theta_1|\bm{d})}\mathrm{d}\theta_1 \mathrm{d}\bm{d} \\
    &\ = \argmin_{\bm{\lambda}} \int \! -P(\theta_1, \theta_2,\bm{d}) \log q_{\bm{\lambda}}(\theta_1|\bm{d})\mathrm{d}\theta_1 \mathrm{d}\theta_2 \mathrm{d}\bm{d} \\
    &\ \approx \argmin_{\bm{\lambda}} \frac{1}{N_{\mathrm{sim}}}\sum_{i=1}^{N_{\mathrm{sim}}} - \log q_{\bm{\lambda}}(\theta_{1,i}|\bm{d}_i) \, .
\end{split}
\end{align}
In the third line we ignored the numerator of the fraction because it has no dependency on $\bm{\lambda}$ and used that $P(\theta_1|\bm{d})P(\bm{d}) = \int \mathrm{d}\theta_2 P(\theta_1, \theta_2, \bm{d})$. The sum introduced in the fourth line runs over joint samples $(\theta_{1,i}, \theta_{2,i}, \bm{d}_{i})$ from  $P(\theta_1, \theta_2, \bm{d})$. Since the summand does not depend on $\theta_{2}$, the $\theta_2$ draws play no role during the optimization of~$\bm{\lambda}$.

This two-parameter example generalizes to cases with many parameters. With SBI, including nuisance parameters is thus straightforward: one defines their prior distribution and includes their effect in the simulations. No architectural changes to the normalizing flow network are needed, provided it has sufficient capacity. When the inferred value of a nuisance parameter is of interest, one may include it in the parameter vector whose distribution the normalizing flow is trained to describe. Otherwise, it may be omitted from the training vector without affecting the marginalized posterior for the parameters of interest.

\begin{figure}[t]
    \centering \includegraphics[width=\columnwidth]{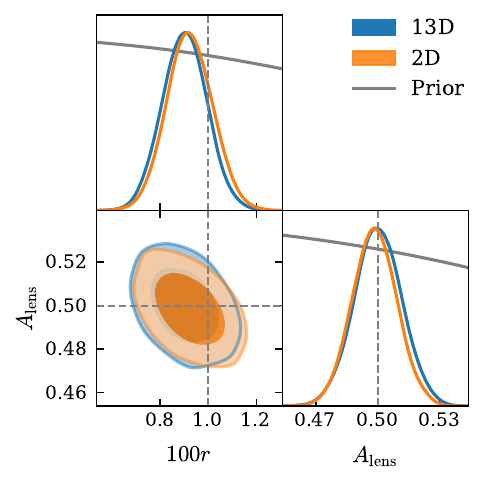}
    \caption{Comparison between posteriors for one of the test datasets described in Sec.~\ref{sec:results_joint_nilc} illustrating that the marginal posterior of $r$ and $A_{\mathrm{lens}}$ is consistent whether the marginal posterior is obtained from the 13-dimensional posterior samples from a normalizing flow that is trained on the 13-dimensional parameter vectors (in blue), or from a normalizing flow trained on the two-dimensional ($r$ and $A_{\mathrm{lens}}$) parameter vector (in orange). }\label{fig:marginal_cosmo_only}
\end{figure}

Fig.~\ref{fig:marginal_cosmo_only} demonstrates the above for the two-dimensional marginal distribution on $r$ and $A_{\mathrm{lens}}$, which is consistent whether the flow is  trained on the full 13-parameter vector or only on $r$ and $A_{\mathrm{lens}}$. We find that the rate at which the NPE method converges with the number of simulations is similar between the two cases. This suggests that it is not the parameter dimensionality, but instead the relatively large dimensionality of the data vector (165) on which the normalizing flow is conditioned that dictates the convergence rate. The small difference in the marginal posterior of $r$ in Fig.~\ref{fig:marginal_cosmo_only} indicates that the normalizing flows have not fully converged.


\bibliography{sample7}{}
\bibliographystyle{aasjournalv7}



\end{document}